\documentclass[structabstract]{aa}

\usepackage{txfonts}
\usepackage{setspace}
\usepackage{natbib}
\bibpunct{(}{)}{;}{a}{}{,} 
\usepackage{graphicx}

\begin{document}

\title{Kinematics of the inner thousand AU region around the young massive star \object{AFGL 2591--VLA3}: a massive disk candidate?}

\author{K.-S. Wang\inst{1}
  \and F. F. S. van der Tak\inst{2,3} 
  \and M. R. Hogerheijde\inst{1}} 

\offprints{Kuo-Song Wang, \email{kswang@strw.leidenuniv.nl}}

\institute{Leiden Observatory, Leiden University, PO Box 9513, 2300 RA Leiden, The Netherlands; email: kswang@strw.leidenuniv.nl 
  \and  SRON Netherlands Institute for Space Research, Landleven 12, 9747 AD Groningen, The Netherlands
  \and Kapteyn Astronomical Institute, University of Groningen, The Netherlands} 

\date{Received/Accepted}

\abstract
{Recent detections of disks around young high-mass stars support the idea of massive star formation through accretion rather than coalescence, but the detailed kinematics in the equatorial region of the disk candidates is not well known, which limits our understanding of the accretion process.}
{This paper explores the kinematics of the gas around a young massive star with millimeter-wave interferometry to improve our understanding of the formation of massive stars though accretion.}
{We use Plateau de Bure interferometric images to probe the environment of the nearby ($\sim$1 kpc) and luminous ($\sim$20000 $L_{\sun}$) high-mass (10--16 $M_{\sun}$) young star \object{AFGL 2591--VLA3} in continuum and in lines of HDO, H$_2$$^{18}$O and SO$_2$ in the 115 and 230 GHz bands. Radiative transfer calculations are employed to investigate the kinematics of the source.}
{At $\sim$0.5$\arcsec$ (500 AU) resolution, the line images clearly resolve the velocity field of the central compact source (diameter of $\sim800$ AU) and show linear velocity gradients in the northeast-southwest direction. Judging from the disk-outflow geometry, the observed velocity gradient results from rotation and radial expansion in the equatorial region of VLA3. Radiative transfer calculations suggest that the velocity field is consistent with sub-Keplerian rotation plus Hubble-law like expansion. The line profiles of the observed molecules suggest a layered structure, with HDO emission arising from the disk mid-plane, H$_2$$^{18}$O from the warm mid-layer, and SO$_2$ from the upper disk.}
{We propose \object{AFGL 2591--VLA3} as a new massive disk candidate, with peculiar kinematics. The rotation of this disk is sub-Keplerian, probably due to magnetic braking, while the stellar wind may be responsible for the expansion of the disk. The expansion motion may also be an indirect evidence of disk accretion in the very inner region because of the conservation of angular momentum. The sub-Keplerian rotation discovered in our work suggests that \object{AFGL 2591--VLA3} may be a special case linking transition of velocity field of massive disks from pure Keplerian rotation to solid-body rotation though definitely more new detections of circumstellar disks around high-mass YSOs are required to examine this hypothesis. Our results support the idea that early B-type stars could be formed with a circumstellar disk from the point of view of the disk-outflow geometry, though the accretion processes in the disk need to be further investigated.}


\titlerunning{\object{AFGL 2591--VLA3}: a massive disk candidate?}
\maketitle 

\section{Introduction}

\begin{table*}[]
	\centering
	\caption{Log of observations}
	\small{
	\begin{tabular}{lcccc}
    \hline \hline
     & 03 Feb 2006\tablefootmark{a} & 19 Feb 2007 & 13 Mar 2007 & 15 Mar 2007 \\
    \hline
    Wavelength & 3\,mm + 1.3\,mm & 3\,mm & 1.3\,mm & 1.3\,mm\\
    On-source time (hours)&	5.3 & 4.5 & 5.0 & 4.0 \\
    Configuration &	6Aq & 6Aq & 6Bq & 6Bq \\
    Bandpass calibrator &	 \object{3C454.3} & \object{3C273} & \object{3C345} & \object{3C273} \\
    Gain calibrator &	  \object{2005+403}, \object{2013+370} & \object{2005+403}, \object{2013+370} & \object{2005+403}, \object{2013+370} & \object{2005+403}, \object{2013+370}\\
    Flux calibrator &	  \object{MWC349} & \object{MWC349} & \object{MWC349} & \object{MWC349} \\
    \hline    
  \end{tabular}
  \tablefoot{ \\
		\tablefoottext{a}{Antenna 4 missing due to tuning problems.} 
		}
		}
	\label{observation_log}	
\end{table*}

Massive stars play an important role in cycling and mixing material between stars and the interstellar medium in galaxies. However, our understanding of how high-mass stars ($M$ $\ge$ 8 $M_{\sun}$) form is still far from complete \citep{Zinnecker07}. Observationally, it is challenging to gain information from high-mass young stellar objects (YSOs) because high-mass stars usually form in groups at large distances (typically a few kpc), requiring high angular resolution to resolve individual sources. High-mass YSOs are embedded in regions with high extinctions ($\sim$10--100 $A_{\rm v}$) for $\sim15\%$ of their lifetime, where complex processes, such as gravitational interaction, powerful outflows and stellar winds, high accretion rates and ionizing radiation fields make the interpretation of observations difficult. Moreover, high-mass YSOs evolve very fast ($\sim10^5$ yrs) and reach the main sequence while still in the embedded phase. Structures such as circumstellar disks have an even shorter lifetime and are harder to detect. 

Accretion via a circumstellar disk is suggested to be the most effective way to overcome radiative pressure and build up a high-mass star \citep[e.g.,][]{Krumholz09}. The detection of massive bipolar outflows with high mass loss rates of $\sim10^{-4}$ $M_{\sun}$ yr$^{-1}$ around young high-mass stars implies the presence of circumstellar disks with comparable mass accretion rates \citep[e.g.,][]{Henning00,Beuther02,Zhang05}. To date, only a dozen or so B-type young stars have been proposed to have circumstellar disks \citep[e.g.,][]{Cesaroni97,Cesaroni99,Cesaroni06,Shepherd99,Beuther05}, with disk masses less than or comparable to the masses of the host stars. In contrast, around young O-type stars, only large rotating toroids are found with masses greater than stellar masses \citep{Beltran05,Furuya08}.

The presence of circumstellar disks around low-mass young stars is firmly established, and our current observational and theoretical understanding is reviewed by \citet{Williams11} and \citet{Armitage11}, respectively. In contrast, the applicability of a scaled-up version of low-mass star formation via disk accretion for massive stars has not been firmly established \citep{Zinnecker07}, although promising cases have been reported \citep[e.g.,][]{Keto10}. Clearly, more observational work is needed to establish the kinematics of material associated with high-mass YSOs.

In this paper, we present observations of massive disk candidate, \object{AFGL 2591--VLA3}, a young early B-type star in the \object{Cygnus X} region. The distance to this object is uncertain, with reported values ranging from 0.5 to 2 kpc \citep[0.5--2 kpc as discussed in][]{vanderTak99} and recent measurements of water masers even suggesting 3.3 kpc \citep{Rygl11}. In this work, we assume a distance of 1 kpc in order to aid comparison with previous publications and discuss how our conclusions change for a distance of 0.5 or 3 kpc. At 1 kpc, the total luminosity of \object{AFGL 2591} is $\sim2\times10^4$ $L_{\sun}$ \citep{Lada84}. Compact radio emission is associated with the source \citep[named as VLA1, VLA2 and VLA3;][]{Campbell84,Trinidad03}. Modeling of its free-free emission implies the stellar mass is about 16 $M_{\sun}$ \citep{vanderTak05}. An envelope mass of only 40 $M_{\sun}$ within 30000 AU around the source is derived through comprehensive radiative transfer modeling of molecular lines by \citet{vanderTak99}.

The geometry of the system can be derived from bipolar $^{12}$CO outflows at arcminute scale, which extend in east (red)-west (blue) direction (position angle (P.A.) $\sim$270$\degr$--325$\degr$) observed at resolutions of $14\arcsec$ to $21\arcsec$ by \citet{Mitchell92} and \citet{Hasegawa95}. Observations of $^{13}$CO at resolution of $14\arcsec$ suggests that the P.A. of the outflow is about $280\degr$ \citep{vanderTak99}, confirmed by $K$-band bispectrum speckle imaging of \object{AFGL 2591} at a resolution of $\sim0.2\arcsec$ showing multiple loops on the western side of the source with P.A. $\sim260\degr$ \citep{Preibisch03}. Precession of the outflow axis is also implied. Dense gas in the outflow walls traced by CS and HCN shows a consistent morphology \citep{Bruderer09}. Based on this orientation of the outflow, a north-south (P.A. $\sim0\degr$) orientation of the equatorial plane of \object{AFGL 2591--VLA3} is inferred. 

Observations of HDO $1_{1,0}$--$1_{1,1}$, H$_2$$^{18}$O $3_{1,3}$--$2_{2,0}$ and SO$_2$ $12_{0,12}$--$11_{1,11}$ with the Plateau de Bure interferometer (PdBI) of the Institut de Radio Astronomie Millim\'etrique (IRAM)\footnote{IRAM is supported by INSU/CNRS (France), MPG (Germany) and IGN (Spain).} reveal a velocity gradient in the  north(east)-south(west) direction across a barely resolved source with a diameter of $\sim800$ AU, suggestive of a circumstellar disk \citep{vanderTak06}. However, the kinematics could not be studied due to the limited angular resolution ($\sim1\arcsec$). Here we present new PdBI observations toward \object{AFGL 2591} at a higher resolution of $\sim0.5\arcsec$ ($\sim500$ AU at 1 kpc), which resolve the kinematics. This paper is organized as follows. Section 2 summarizes the PdBI observations and data reduction. Section 3 presents the results. In Sect. 4, we describe a simple model of the kinematics. Discussions and conclusions are given in Sect. 5 and Sect. 6, respectively. 

\section{Observations}
	The source \object{AFGL 2591} was observed with PdBI in 2006--2007 in 4 tracks. The 3 mm receiver was tuned to cover HDO 1$_{1,0}$--1$_{1,1}$ at 80578.3 MHz (hereafter HDO) and the 1.3 mm receiver was tuned to cover H$_2$$^{18}$O $3_{1,3}$--$2_{2,0}$ at 203407.5 MHz (hereafter H$_2$$^{18}$O) and SO$_2$ $12_{0,12}$--$11_{1,11}$ at 203391.6 MHz (hereafter SO$_2$). Table \ref{observation_log} presents a log of the observations and Table \ref{mol_line_table} summarizes the observed lines. The combined projected baselines range from 26 k$\lambda$ ($\sim$10$\arcsec$) to 206 k$\lambda$ in length at 3 mm and from 48 k$\lambda$ ($\sim$5$\arcsec$) to 525 k$\lambda$ in length at 1.3 mm. The phase tracking center was chosen at $\alpha(2000)$ = 20$^{\rm h}$29$^{\rm m}$24$\fs$87, $\delta(2000)$ = 40$\degr$11$\arcmin$19$\farcs$50. The nominal LSR velocity adopted for observations was $-5.5$ km s$^{-1}$. The quasars \object{2005+403} and \object{2013+370}, which are about 5$\degr$ away from \object{AFGL 2591}, were frequently observed for phase and gain calibrations in all four observations. \object{3C273}, \object{3C345} and \object{3C454.3} were used as the bandpass calibrators, and \object{MWC349} with a total flux density of $\sim$1.3 Jy at 203.4 GHz during the time of the observations, as flux calibrator. The flux error is estimated to be within 20\%. The spectral resolutions are 39.0625 kHz per channel or 0.145 km s$^{-1}$ per channel at 3 mm band, and 78.125 kHz per channel or 0.115 km s$^{-1}$ per channel at 1.3 mm band. Data reduction was conducted at the IRAM headquarters in Grenoble, using the GILDAS\footnote{http://www.iram.fr/IRAMFR/GILDAS} software. 

\begin{table}[htbp]
	\centering
	\caption{Observed molecular lines}	
	\small{
	\begin{tabular}{lcccc}
		\hline\hline
		  &  & $\nu$\tablefootmark{a} & $E_{\rm u}$\tablefootmark{b} & Beam Size\tablefootmark{c} \\
		  Molecule & Transition & (MHz) & (K) & ($\arcsec\times\arcsec; \degr$) \\ 	
		\hline
		HDO	&	$1_{1,0}$--$1_{1,1}$	&	80578.3  & 47 & $1.23\times0.72$; 26 \\
		H$_2$$^{18}$O &	$3_{1,3}$--$2_{2,0}$	&  203407.5  & 204 & $0.51\times0.33$; 22 \\
		SO$_2$ &		$12_{0,12}$--$11_{1,11}$	&  203391.6  & 70 & $0.51\times0.33$; 22 \\
		\hline
	\end{tabular}
	}					
	\tablefoot{ \\
		\tablefoottext{a}{Rest frequency taken from JPL molecular spectroscopy (http://spec.jpl.nasa.gov/).} \\
		\tablefoottext{b}{Energy of upper level.} \\
		\tablefoottext{c}{Synthesized beam using natural weighting.}}	
	\label{mol_line_table}
\end{table}

\section{Results}
\begin{figure}[h]
	\centering
	\includegraphics[width=9cm]{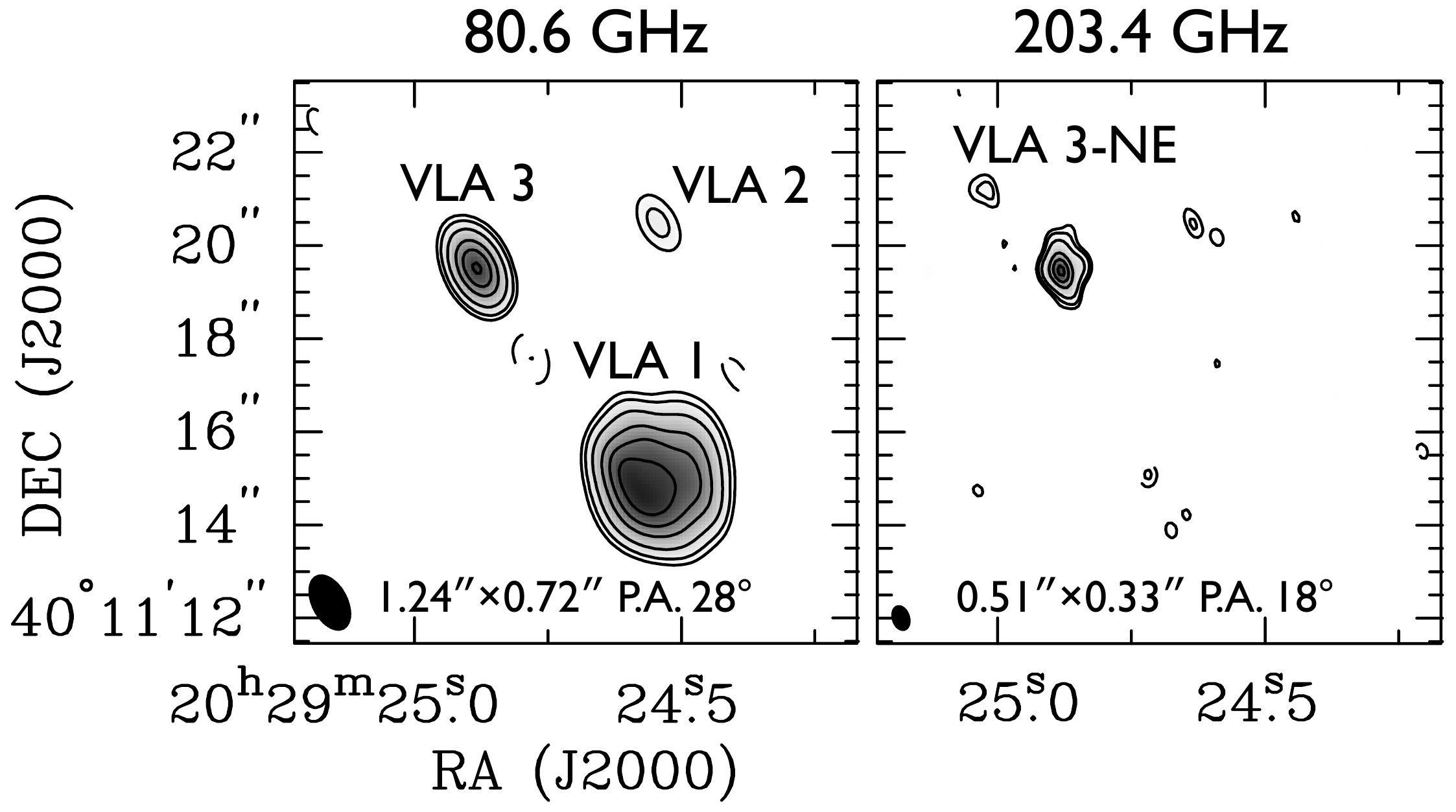}
	\caption{Natural-weighted continuum images in the region of \object{AFGL 2591}. (Left) 80.6-GHz continuum. Contours are at 3, 5, 10, 20, 30... $\sigma$ with 1-$\sigma$ of 0.15 mJy beam$^{-1}$. Sources within the field following \citet{Campbell84} and \citet{Trinidad03} are marked on the map. (Right) 203.4-GHz continuum. Contours are at 3, 5, 10, 30, 50... $\sigma$ with 1-$\sigma$ of 0.4 mJy beam$^{-1}$.}
	\label{continuum_maps}
\end{figure}

\begin{figure}[h]
	\centering
	\includegraphics[width=9cm]{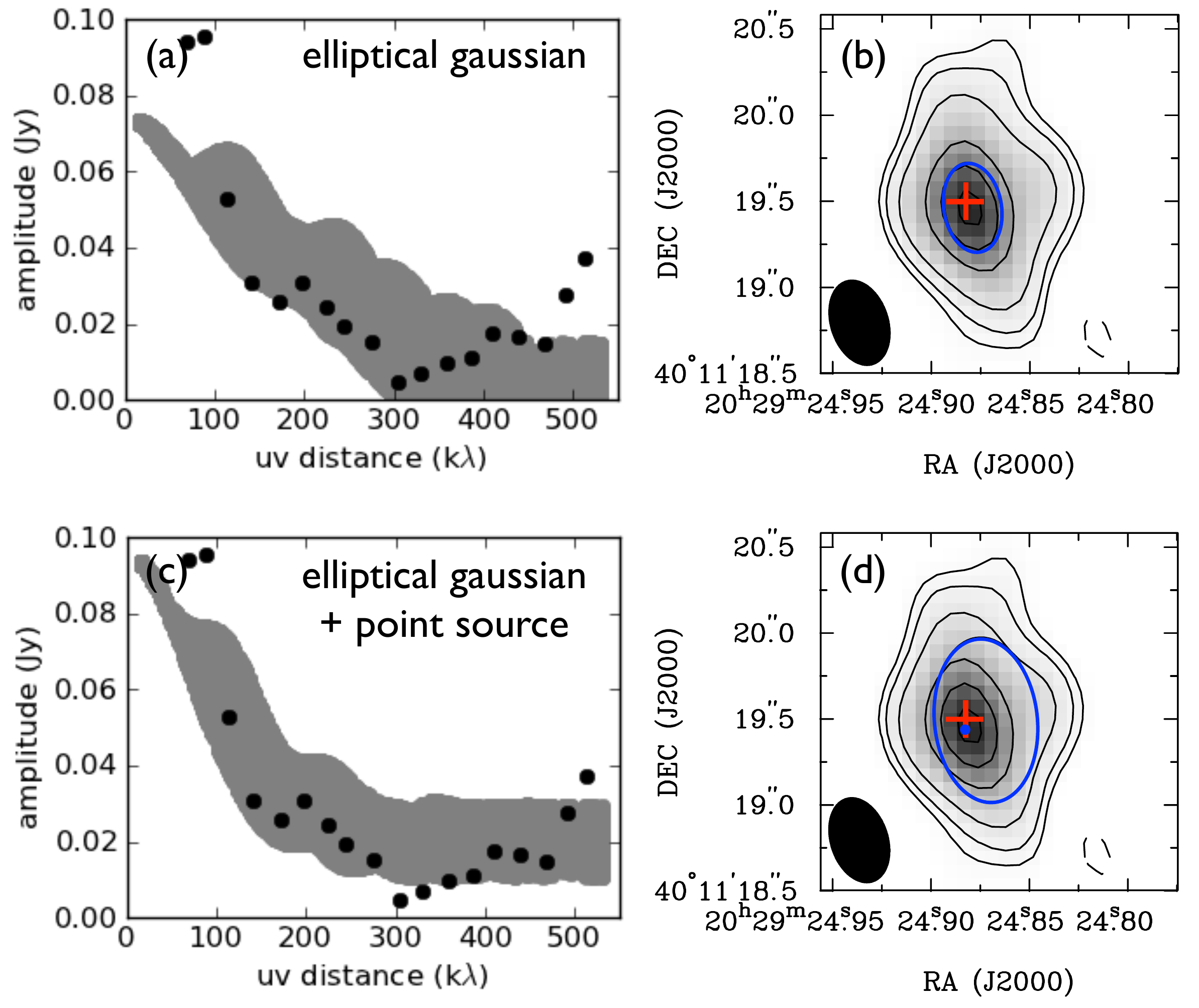}
	\caption{Visibility analysis of the continuum at 203.4 GHz. (a) and (c) show the observed visibilities (filled circles) and model visibilities (grey area). The observed data are vector-averaged for the purpose of better visualization. Only the averaged structure in circular symmetry can be observed in the plot. Full model visibilities are plotted in order to show the non-circular symmetric structure of the source. (b) and (d) represent a graphical view of the visibility fits of VLA3 shown in (a) and (c), respectively. The red plus sign is the peak position of the emission. Blue ellipse and filled circle stand for the Gaussian component and the point source.}
	\label{cont_vis_analysis}
\end{figure}		

	\subsection{Continuum emission}
		The natural-weighted continuum images of \object{AFGL 2591--VLA3} at 80.6 GHz and 203.4 GHz are presented in Fig. \ref{continuum_maps}. The angular resolutions are $1\farcs24\times0\farcs72$, P.A. $28\degr$ at 80.6 GHz and $0\farcs51\times0\farcs33$, P.A. $18\degr$ at 203.4 GHz. With an arcsecond beam, VLA3 is unresolved at 80.6 GHz. In contrast, at 203.4 GHz VLA3 is well resolved with a subarcsecond beam and shows a diamond shaped morphology in the north-south direction. In addition to VLA3, VLA1 and VLA2 are detected within the field of view at 80.6 GHz. At 203.4 GHz, VLA1 is likely resolved out due to missing  short spacings. A previously unknown source (VLA3--NE) located at $\sim$2.5$\arcsec$ north-eastern to VLA3 is detected at 5-$\sigma$ at 203.4 GHz. 

	\subsubsection{Measurements of positions, sizes and total flux densities}			
	The positions, sizes and total flux densities of the sources in the region of \object{AFGL 2591} are derived in the visibility domain and summarized in Table \ref{continuum_parameter}. Judging from the images in Fig. \ref{continuum_maps}, all sources are fit with point sources, except VLA1 at 80.6 GHz and VLA3 at 203.4 GHz, where Gaussians are used. The vector-averaged visibility plot in Fig. \ref{cont_vis_analysis} shows a compact component on baselines longer than 200 k$\lambda$ for VLA3 at 203.4 GHz. We therefore fit VLA3 with a 20 mJy point source plus a 65 mJy, $1.0\arcsec\times0.6\arcsec$ ($\sim800$ AU diameter at 1 kpc) Gaussian source. Including the flux densities of VLA2 and VLA3--NE, the total flux density within the field of view at 203.4 GHz is about 93 mJy which is consistent with the visibility measurement (Fig. \ref{cont_vis_analysis}).
	
		Compared to \citet{vanderTak06}, our 80.6-GHz observations measure  100\%, 12\% and 45\% of the total flux densities of VLA1, VLA2 and VLA3, respectively, while the new 203.4-GHz observations detect 0\%, 7\% and 44\% of the total flux densities of VLA1, VLA2 and VLA3, respectively. Considering the different array configurations (limited by the antenna shadowing effect, the shortest baselines of the observations made by \citet{vanderTak06} are 8 k$\lambda$ at 3 mm and 20 k$\lambda$ at 1.3 mm) and the absolute uncertainty of the flux density, the new data are consistent with optically thin free-free emission, optically thick dust or free-free emission and optically thin dust emission for VLA1, VLA2 and VLA3, respectively, as discussed by \citet{Trinidad03} and \citet{vanderTak99,vanderTak06}. For the newly detected source VLA3--NE, more sensitive high-resolution observations at different frequencies are needed to derive its nature.

\begin{table*}
	\centering
	\caption{Characteristics of Continuum Emission}
	\small{	
	\begin{tabular}{lccccccc}
		\hline\hline
			&	$\Delta\alpha$\tablefootmark{a}	&	$\Delta\delta$\tablefootmark{a}	&	$\theta_{\rm major}$\tablefootmark{b}	&	$\theta_{\rm minor}$\tablefootmark{b}	&	P.A.\tablefootmark{b}  & $F_{\nu}$\tablefootmark{c}  & $F_{\nu}$(vdT06)\tablefootmark{d} \\
		Source	&	($\arcsec$)	&	($\arcsec$)	&	($\arcsec$)	&	($\arcsec$)	&	($\degr$) 	& (mJy) 	& (mJy)	\\
		\hline			
		\multicolumn{8}{c}{80.6 GHz} \\
		\hline			
		VLA1	&   $-3.67$	&   $-4.49$	&   2.28	&   2.13	&   $-18\pm7$	&   $62\pm1$		&   $61\pm1$   \\
		VLA2	&   $-3.64$	&   1.31	&   ...	&   ...	&   ...			&   $1.1\pm0.1$		&   $9\pm1$	    \\
		VLA3		&   0.11	&   $-0.04$	&   ...	&	...	&   ...			&	$7.2\pm0.1$		&   $16\pm2$	\\		
		\hline
		\multicolumn{8}{c}{203.4 GHz} \\
		\hline
		VLA1	&   ...			&   ...			&   ...			&   ...			&   ...			&   ...			&   $61\pm2$   \\
		VLA2	&   $-2.75$	&   0.94	&   ...			&	...			&   ...			&   $4.7\pm0.4$		&   $65\pm4$   \\
		VLA3	\tablefootmark{e}		&   0.03		&   0.00	&   0.96	&   0.60	&   $6\pm2$		&   $65\pm2$		&   $194\pm1$  \\
									&   0.16		&   $-0.05$	&   ...			&   ...			&   ...			&   $20\pm1$		&	...		\\
		VLA3--NE	&   	1.80	&   1.82	&   ...			&	...			&   ...			&   $4.7\pm0.4$		&	...		\\		
		\hline			
		\end{tabular}
	\tablefoot{
		Positions, source sizes and total flux densities are derived in visibility domain.\\
		\tablefoottext{a}{Position offset with respect to $\alpha(2000) = 20^{\rm h}29^{\rm m}24\fs87$ and $\delta(2000) = 40\degr11\arcmin19\farcs50$.} \\
		\tablefoottext{b}{FWHM source size and position angle.} \\
		\tablefoottext{c}{Total flux density. The absolute flux uncertainty is $\sim20\%$. } \\
		\tablefoottext{d}{Total flux density measured by \citet{vanderTak06} in the visibility domain.} \\
		\tablefoottext{e}{Model with a Gaussian plus a point source. See Fig. \ref{cont_vis_analysis} and text for more discussions.} 
	}
	}
	\label{continuum_parameter}
\end{table*}
		
	\subsubsection{Estimation of densities and masses}	
		For the ``dust'' source VLA3, we derive the beam averaged H$_2$ column density from the measured peak intensity of 30 mJy per beam at 203.4 GHz. The H$_2$ column density at the emission peak can be estimated as 
	\begin{equation}
		N({\rm H_2}) = \frac{S_\nu a}{2 m_{\rm H} \Omega_{\rm b} \kappa_\nu B_\nu(T_{\rm d})},
	\end{equation}
where $S_\nu$ is the peak flux density, $a$ is the gas-to-dust ratio (100), $\Omega_{\rm b}$ is the beam solid angle, $m_{\rm H}$ is the mass of atomic hydrogen, $\kappa_\nu$ is the dust opacity per unit mass and $B_\nu(T_{\rm d})$ is the Planck function at dust temperature $T_{\rm d}$. We apply the value of $\kappa_\nu(1.3 {\rm mm}) \approx 1.0$ cm$^2$ g$^{-1}$ as suggested by \citet{Ossenkopf94} for gas densities of $10^6-10^8$ cm$^{-3}$ and coagulated dust particles with thin ice mantles. The beam averaged H$_2$ column densities for $T_{\rm d}$ from 100 to 300 K (see Sect. 3.2.4 for the validity of this assumption) are $0.5-1.7\times10^{24}$ cm$^{-2}$. The gas mass is estimated from the integrated continuum flux density of 85 mJy via
	\begin{equation} 
		M_{\rm gas} = \frac{F_\nu d^2 a}{\kappa_\nu B_\nu(T_{\rm d})},
	\end{equation}
where $F_\nu$ is the total flux density of dust emission and $d$ is the distance to the source. At a distance of 1 kpc and temperatures of 100--300 K, the mass is 0.1--0.3 $M_{\sun}$, significantly less than the stellar mass of VLA3 \citep[$\approx$10--16 $M_{\sun}$,][]{Lada84, vanderTak05}. If the distance is 3 kpc, the gas mass is modified to 1--3 $M_{\sun}$, still less than the scaled stellar mass of $>20$ $M_{\sun}$. Regardless of the distance, the inferred mass of the surrounding gas around VLA3 is much less than the stellar mass. With the further assumption of a spherical source of $0.8\arcsec$ in diameter, as measured from the visibilities of VLA3 at 203.4 GHz, H$_2$ number densities of $7-22\times10^{7}$ cm$^{-3}$ are derived for a distance of 1 kpc, and of $2-7\times10^{7}$ cm$^{-3}$ for a distance of 3 kpc. These densities are lower limits if the source structure is flattened. The inferred gas number density is consistent with the envelope density profile within radii of 400 AU ($10^{7-9}$ cm$^{-3}$) as derived by \citet{vanderTak99}. We summarize these results in Table \ref{disk_mass_density}.

\begin{table}
	\centering
	\caption{H$_2$ column density, mass and number density of VLA3}
	\small{
	\begin{tabular}{lcccc}
    \hline \hline
    		&	 \multicolumn{3}{c}{Dust Temperature} \\
		\cline{2-4}
    		&	 100 K	&	200 K	& 300 K  & Note \\
    \hline
    $N({\rm H_2})\ / \ 10^{24}$ cm$^{-2}$ & 1.7 & 0.8 & 0.5 \\
    \hline
    $M_{\rm gas}$\ / \ $M_{\sun}$ & 0.34 & 0.16 & 0.11 & $d=1$ kpc \\
    $n({\rm H_2})\ / \ 10^{7}$ cm$^{-3}$ & 22.3 & 10.8 & 7.2 & if size = 800 AU \\
    \hline
    $M_{\rm gas}$\ / \ $M_{\sun}$ & 3.03 & 1.48 & 0.98 & $d=3$ kpc \\
    $n({\rm H_2})\ / \ 10^{7}$ cm$^{-3}$ & 7.5 & 3.6 & 2.4 &  if size = 2400 AU \\
    \hline    
    \end{tabular}
	}
	\label{disk_mass_density}
\end{table}
\begin{table}
	\caption{Source positions and size parameters of VLA3 derived from molecular species.}	
	\small{
	\begin{tabular}{lcccccc}
		\hline\hline
		    &   $\Delta\alpha$\tablefootmark{a}	&	$\Delta\delta$\tablefootmark{a}	&	$\theta_{\rm major}$\tablefootmark{b}	&	$\theta_{\rm minor}$\tablefootmark{b}	&	P.A.\tablefootmark{b}  &  $V_{\rm grad}$ P.A.\tablefootmark{c} \\
		Molecule  &  ($\arcsec$)	&	($\arcsec$)	&	($\arcsec$)	&	($\arcsec$)	&	($\degr$) & ($\degr$) \\    	
		\hline
		HDO  &  0.13  &  $-0.08$  &  1.03  &  0.86  &  $48\pm15$  &  $54\pm12$  \\
		H$_2$$^{18}$O\tablefootmark{d}  &  0.07  &  0.05  &  0.84  &  0.39  &  $15\pm7$  &  $40\pm7$ \\
		                                &  0.14  &  0.06  &  ...   &  ...   &  ...       &  ...  \\
		SO$_2$\tablefootmark{d}  &  0.09  &  $-0.08$  &  0.98  &  0.92  &  $-2\pm5$  &  $33\pm9$ \\
		                         &  0.13  &  0.12   &  ...  & ...  &  ...  & ... \\
		\hline
	\end{tabular}
	}					
	\tablefoot{Positions and source sizes are derived in visibility domain.\\
		\tablefoottext{a}{Position offset with respect to $\alpha(2000) = 20^{\rm h}29^{\rm m}24\fs87$ and $\delta(2000) = 40\degr11\arcmin19\farcs50$.} \\
		\tablefoottext{b}{FWHM source size and position angle.} \\
		\tablefoottext{c}{Position angle of the velocity gradient measured from blue- and red-shifted components. See also Fig. \ref{line_maps}.} \\
		\tablefoottext{d}{Model with a Gaussian plus a point source.}}	
\label{tab:mol_size}
\end{table}	

	\subsection{Line emission}
\begin{figure*}[htbp]
	\centering
	\includegraphics[width=17cm]{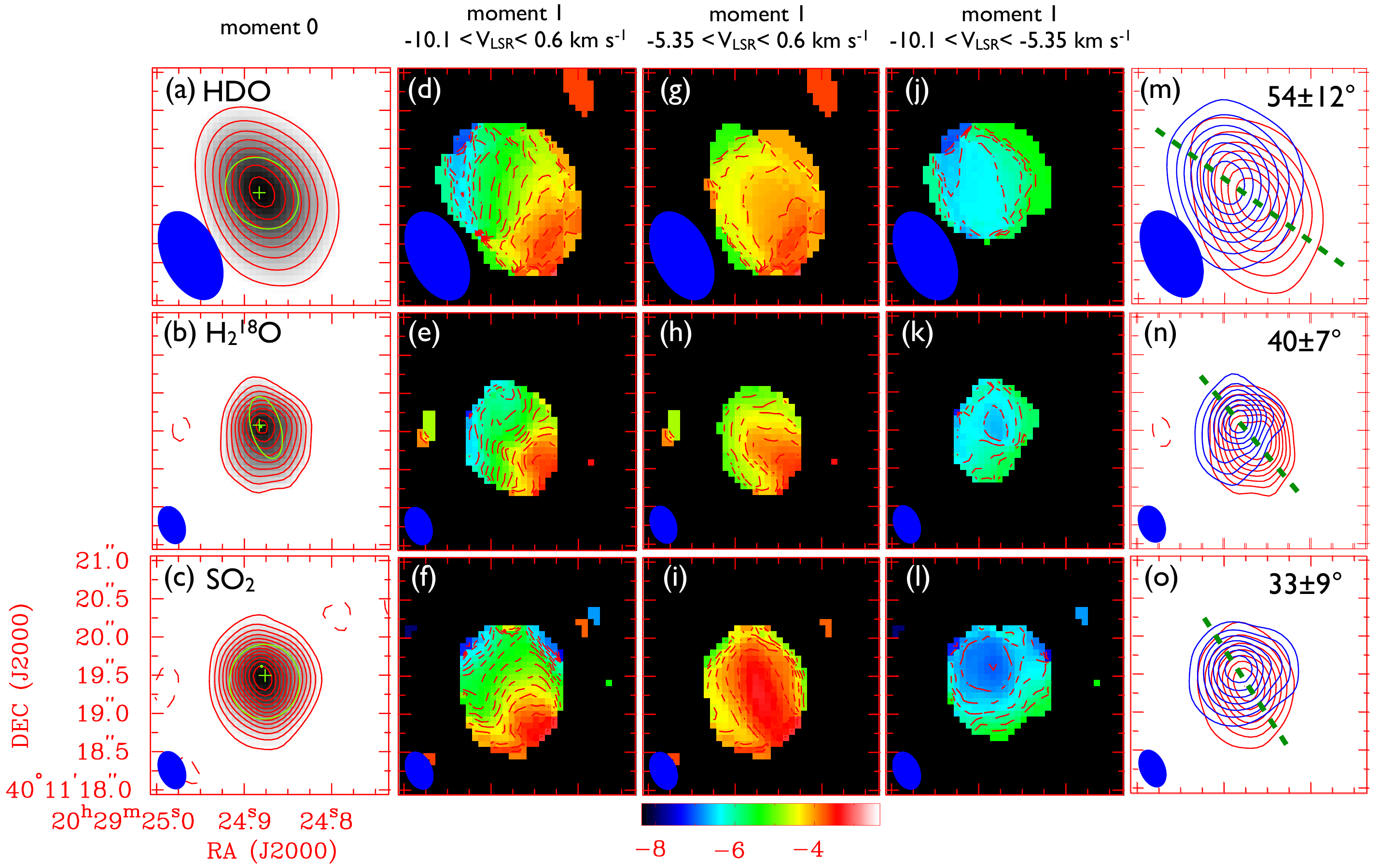}
	\caption{(a)--(c) Zero moment maps of HDO, H$_2$$^{18}$O and SO$_2$ around VLA3, respectively. The contours are at $-3$, 3, 6, 9...$\sigma$ for all three maps, where $1\sigma$ noise levels are 10, 37 and 65 mJy km s$^{-1}$ for HDO, H$_2$$^{18}$O and SO$_2$, respectively. The green crosses mark the positions of emission peaks. The green ellipses represent FWHM sizes derived from visibility fit. The green filled circles indicate the positions of the additional point sources in the visibility fit of H$_2$$^{18}$O and SO$_2$. The natural-weighted clean beam sizes, shown in blue filled ellipses, of HDO and H$_2$$^{18}$O/SO$_2$ are $1\farcs23\times0\farcs72$, P.A. 26$\degr$ and $0\farcs51\times0\farcs33$, P.A. 22$\degr$, respectively. (d)--(l) First moment maps of HDO, H$_2$$^{18}$O and SO$_2$ around VLA3, respectively. For all nine maps, the color scales are set to be the same. Dashed lines represent the isovelocity contours with an interval of 0.25 km s$^{-1}$. (m)--(o) Molecular emission integrated over the velocity ranges of $-5.35 < V_{\rm LSR} < 0.6$ km s$^{-1}$ (red contours) and $-10.1 < V_{\rm LSR} < -5.35$ km s$^{-1}$ (blue contours) for HDO, H$_2$$^{18}$O and SO$_2$, respectively. Contours are plotted with $10\%$ interval of peak value of each velocity component. The inferred position angles of the velocity gradients across VLA3 are plotted as green dashed lines.}
	\label{line_maps}
\end{figure*}

		Compact emission of HDO, H$_2$$^{18}$O and SO$_2$ is found toward VLA3 (Fig. \ref{line_maps}(a)--(c)). With the improved angular resolutions of our new observations (natural weighting; 3mm: $1\farcs23\times0\farcs72$, P.A. 26$\degr$; 1.3mm: $0\farcs51\times0\farcs33$, P.A. 22$\degr$), all three molecular images are resolved and show a nearly round shape, unlike the diamond shape with a centrally peaked feature seen in the 203.4-GHz continuum. At $1\sigma$ noise level, no significant line emission is found toward other sources in \object{AFGL 2591}.

		\subsubsection{Measurements of positions and sizes}	
\begin{figure}[t]
	\centering
	\includegraphics[width=5.5cm]{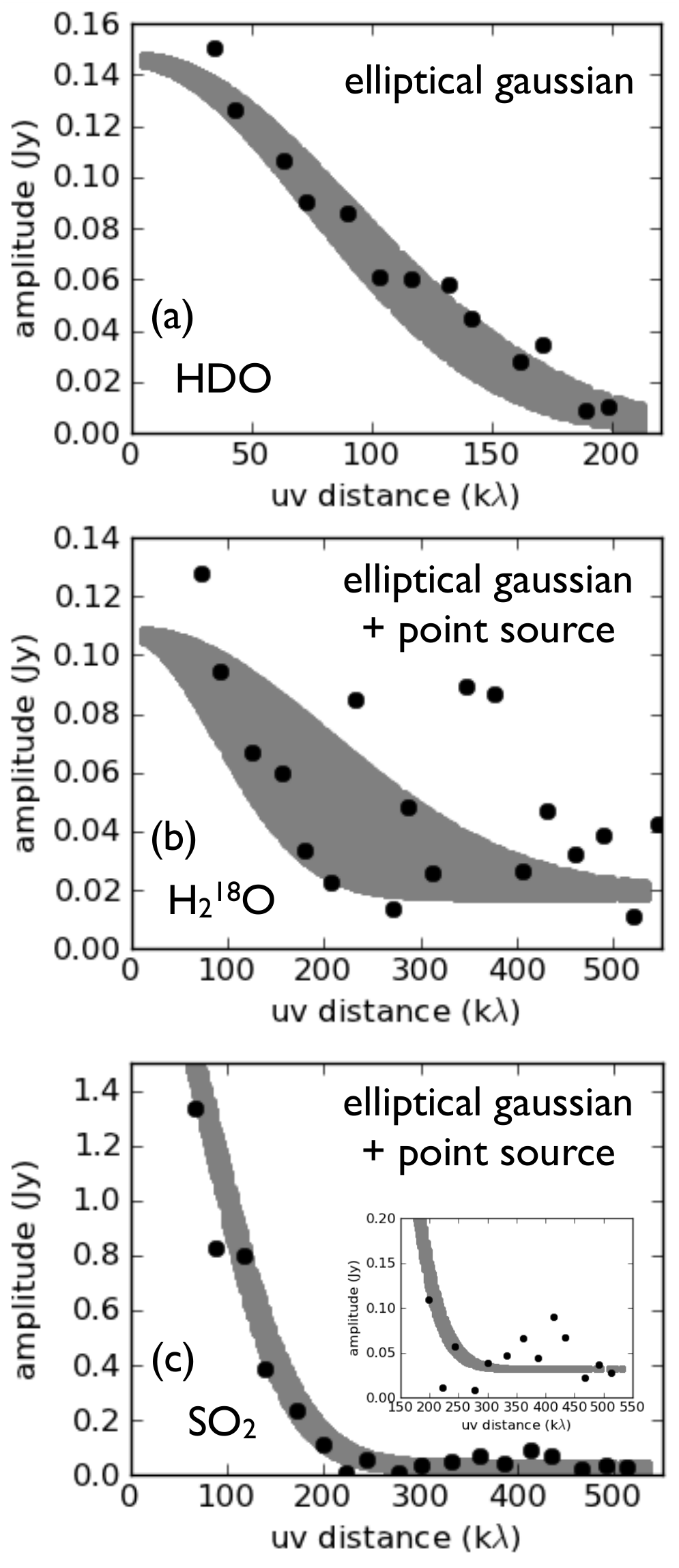}
	\caption{Observed visibilities of HDO, H$_2$$^{18}$O and SO$_2$ (filled circles) overplot with model visibilities (grey area). The observed data are vector-averaged corresponding to the circularly symmetric structure. Full model visibilities are plotted in order to show the non-circular symmetric structure of the source.}
	\label{line_vis_plot}
\end{figure}

			We derive the positions and the sizes of the line emission in the visibility domain (Fig. \ref{line_vis_plot}). In the vector-averaged visibility plots, the amplitude of the visibility decreases up to 200 k$\lambda$ and is more or less flat onward. This feature is similar to what we observed in the continuum visibility at 203.4 GHz, which implies a point source surrounded by an extended source. As a result, we adopted an elliptical Gaussian plus a point source in the visibility fit of H$_2$$^{18}$O and SO$_2$. For HDO, we used a Gaussian only in the fit since our baselines at 80.6 GHz only extend to 200 k$\lambda$. The source sizes derived in this approach are 0.4$\arcsec$--1.0$\arcsec$. The lower end of the result is estimated from H$_2$$^{18}$O, which could be due to the large dispersion of the visibilities or reflect a smaller source size traced by this higher excitation line ($E_{\rm u}=204$ K; 47 and 70 K for HDO and SO$_2$, respectively). The size measurements of the molecular lines are consistent with those derived from the continuum, indicating that the dust and the molecular gas have a similar spatial distribution. The locations of the emission peaks differ by much less than the synthesized beam and are negligible.  We summarize the results of the visibility analysis in Table \ref{tab:mol_size}.			
		
		\subsubsection{Velocity gradients across VLA3}
			The new dataset allows us to spatially resolve the velocity distribution across VLA3. In Fig. \ref{line_maps} (d)--(f), velocity gradients in the northeast-southwest direction can be seen with the red-shifted part toward the southwest side and the blue-shifted part toward the northeast side. Apart from the difference in angular resolution, HDO and H$_2$$^{18}$O show a  similar distribution in isovelocity contours. Detailed investigation of H$_2$$^{18}$O image reveals that the isovelocity contours near the systemic velocity (green part) exhibits a mirrored S shape, while SO$_2$ shows a normal S-shaped distribution of isovelocity contours. The difference implies that HDO/H$_2$$^{18}$O and SO$_2$ trace different kinematics of VLA3, where SO$_2$ may be dominated by the east-west outflow. First moment maps plotted in the velocity ranges of $-5.35 < V_{\rm LSR} < 0.6$ km s$^{-1}$ (Fig. \ref{line_maps}(g)--(i)) and $-10.1 < V_{\rm LSR} < -5.35$ km s$^{-1}$ (Fig. \ref{line_maps}(j)--(l)) support our speculation that SO$_2$ emission is dominated by the outflow since the extreme velocities originate near the source center, while HDO and H$_2$$^{18}$O are less affected by the outflow and presumably dominated by the equatorial motion. We derive the position angles of the velocity gradients across the source seen in Fig. \ref{line_maps} (d)--(f) by integrating over the velocity ranges of $-5.35 < V_{\rm LSR} < 0.6$ km s$^{-1}$ and $-10.1 < V_{\rm LSR} < -5.35$ km s$^{-1}$ for each line and overplotting them in Fig. \ref{line_maps} (m)--(o), respectively. The position angles of velocity gradients are calculated from the emission peaks of the blue- and red-shifted components derived from Gaussian fit. For HDO, H$_2$$^{18}$O and SO$_2$, the position angles are $54\pm12\degr$, $40\pm7\degr$ and $33\pm9\degr$, respectively (Table \ref{tab:mol_size}).

			The orientation of the observed velocity gradients (northeast-southwest) is not perpendicular to the large scale outflow (east-west) \citep{Mitchell92,Hasegawa95,vanderTak99} with a difference of $\sim45\degr$ in position angle. We rule out the possibility of contamination from the bipolar outflow since the red-shifted emission of the molecular emission lies to the southwest, which is inconsistent with the red lobe of the outflow in the east. High-resolution $K$-band imaging \citep{Preibisch03} shows that the orientation of the large-scale outflow remains unchanged on small scales, ruling out contamination by a small-scale outflow to explain the observed velocity pattern.  A more likely interpretation for the observed velocity pattern is rotation in the equatorial plane around \object{AFGL 2591--VLA3}. For purely azimuthal motion we would expect the velocity gradient to be perpendicular to the bipolar outflow direction. Section 4 further explores the effect of radial motion in the equatorial plane that affects the orientation of the velocity gradient.  
						
\begin{figure*}[t]
	\centering
	\includegraphics[width=13cm]{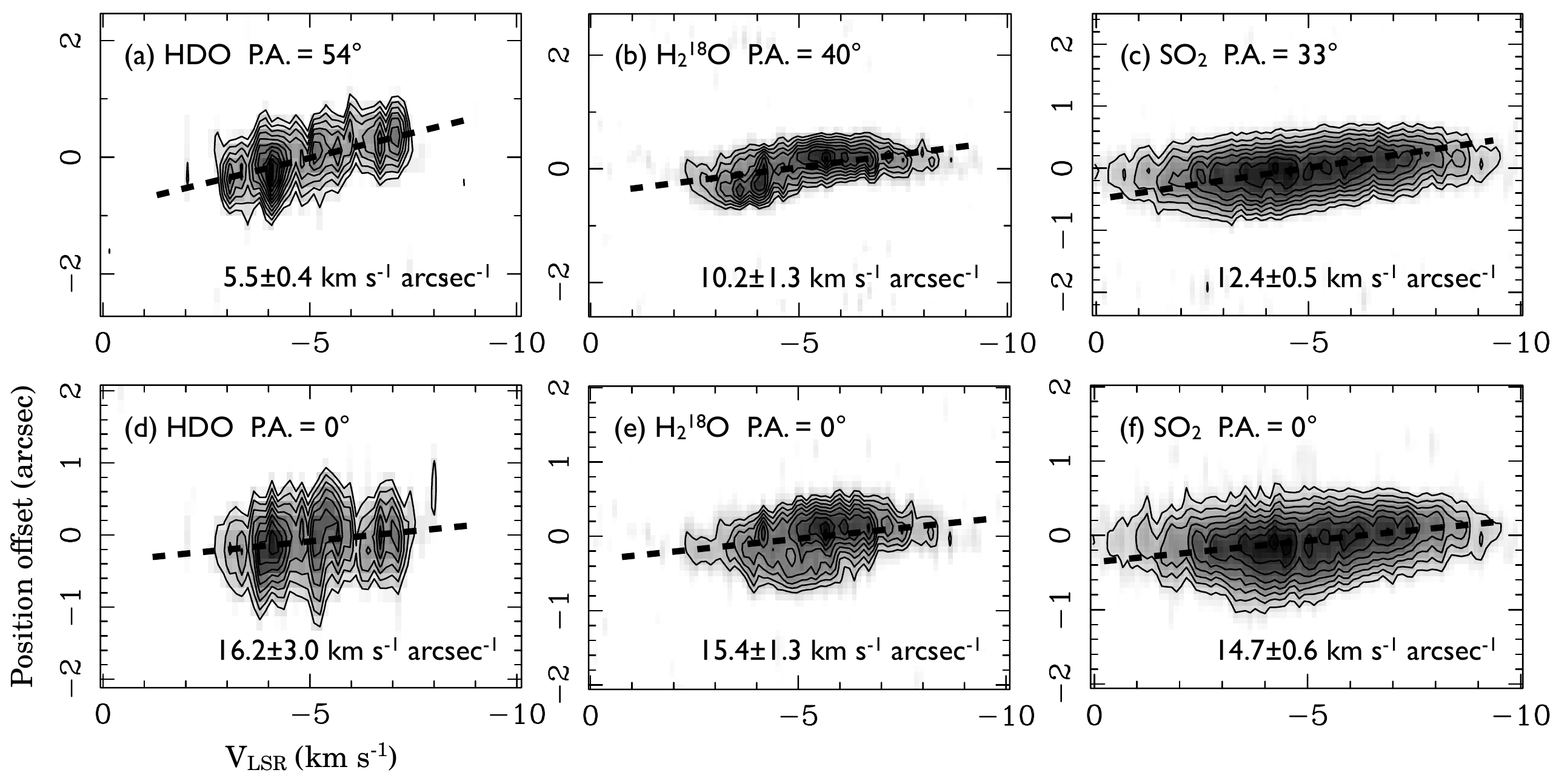}
	\caption{(a)--(c) Position-Velocity maps of HDO (P.A. $54\degr$; see Fig. \ref{line_maps} (m)--(o)), H$_2$$^{18}$O (P.A. $40\degr$) and SO$_2$ (P.A. $33\degr$). Linear velocity gradients are fitted to the maps and shown as dotted lines. (d)--(f) Same as (a)--(c) except the position angle of the position-velocity cut for all three molecules is $0\degr$.}
	\label{fig:pvmaps}
\end{figure*}

			To further quantify the observed velocity gradients, we plot the position-velocity maps with the position angles defined in Fig. \ref{line_maps} (m)--(o) (Fig. \ref{fig:pvmaps}). Clearly, in all cases a $linear$ velocity gradient is found, which is very different from the typical Keplerian rotation found in the protoplanetary disks around low-mass stars \citep[e.g.,][]{Simon00}. We derive (projected) velocity gradients of $5.5\pm0.4$, $10.2\pm1.3$ and $12.4\pm0.5$ km s$^{-1}$ arcsec$^{-1}$ or km s$^{-1}$  per 1000 AU (at 1 kpc) for HDO, H$_2$$^{18}$O and SO$_2$, respectively, suggesting that SO$_2$ is more affected by the east-west outflow while HDO is less affected. To minimize the contamination of the outflow and highlight the motion in the equatorial plane of VLA3, we also derive the velocity gradients by taking the position-velocity cut with P.A. = $0\degr$. A consistent strength of linear velocity gradient ($\sim15$ km s$^{-1}$ arcsec$^{-1}$) is derived from all three molecules, indicating that the effect of outflow is indeed minimized.  		
		
		\subsubsection{Molecular line profile}			
\begin{figure}[htbp]
	\centering
	\includegraphics[width=5cm]{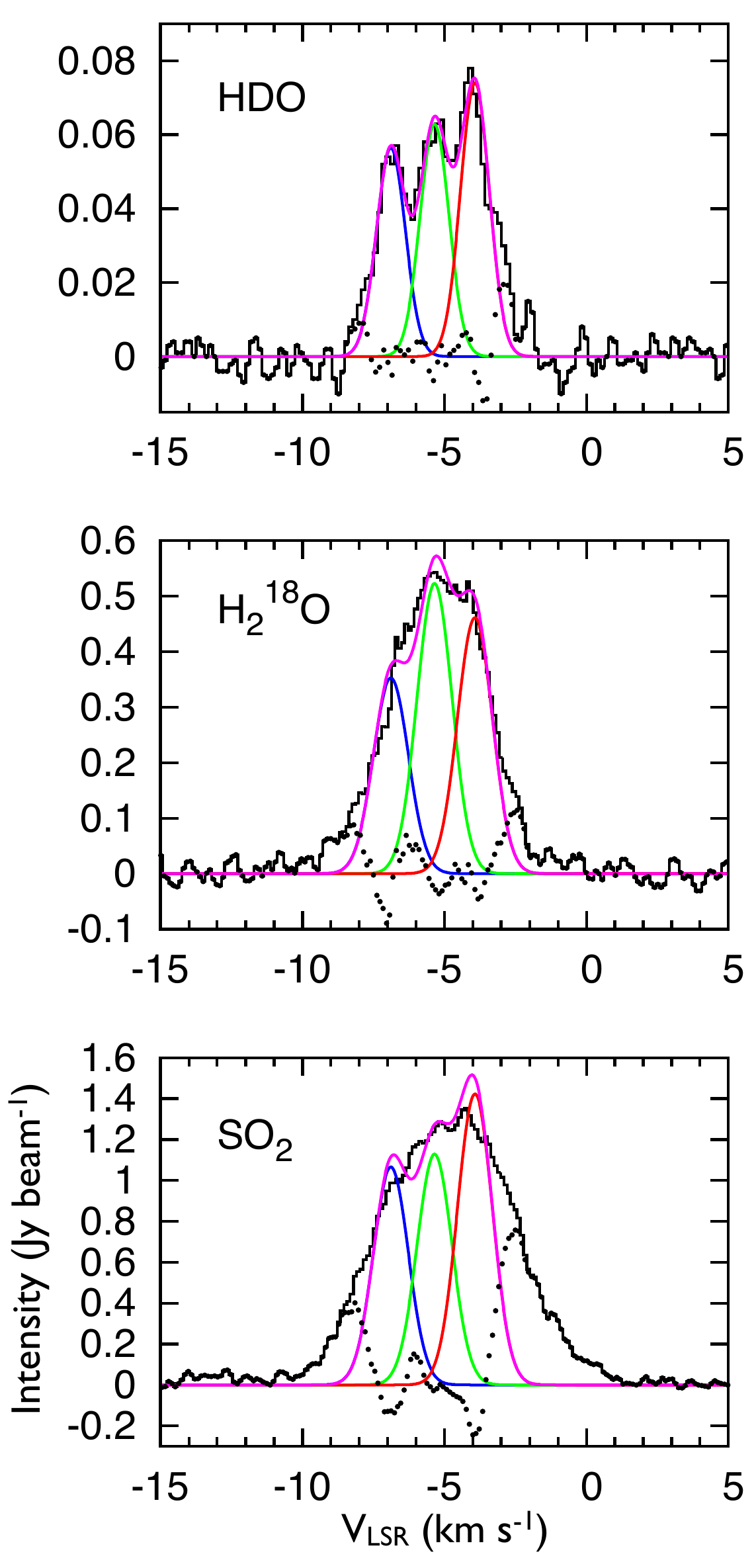}
	\caption{Molecular spectra taken at the mean emission peak after convolving all data to the same angular resolution of $1\farcs23\times0\farcs72$, P.A. 26$\degr$. Fitted Gaussians are overplotted in color. Residuals are shown in dots.}
	\label{fig:peak_spec}
\end{figure}

			Spectra of HDO, H$_2$$^{18}$O and SO$_2$ centered at the mean emission peak ($\Delta\alpha=-0.15\arcsec$, $\Delta\delta=0.02\arcsec$) in a single synthesized beam ($1\farcs23\times0\farcs72$, P.A. 26$\degr$) are shown in Fig. \ref{fig:peak_spec}. A triple-peaked line profile is observed for HDO, while the H$_2$$^{18}$O and SO$_2$ line profiles are smooth Gaussians with weak hints of secondary peaks coincident with those  seen in HDO. We use three Gaussians to decompose the HDO line profile (Table \ref{tab:lines}) and find that Gaussians with FWHM of $\sim1.20\pm0.05$ km s$^{-1}$ at $V_{\rm LSR}$ of $-6.89\pm0.03$, $-5.35\pm0.34$ and $-3.92\pm0.26$ km s$^{-1}$ best reproduce the observed spectrum, although significant residuals remain near $-8.0$ and $-3.0$ km s$^{-1}$. The velocity differences between the blue-shifted and systemic components ($1.54\pm0.34$ km s$^{-1}$), and, the red-shifted and systemic components ($1.43\pm0.43$ km s$^{-1}$) are comparable, which suggest that the red- and blue-shifted components are likely from the same structure undergoing systematic motion, not just a spatial coincidence of three different gas clumps. To decompose the H$_2$$^{18}$O and SO$_2$ line profiles, we fit three Gaussians fixed at $V_{\rm LSR}$ of $-6.89$, $-5.35$ and $-3.92$ km s$^{-1}$ derived from HDO line profile (Table \ref{tab:lines}). For an FWHM line width of 1.45 km s$^{-1}$, this fit reproduces most of the line profile, with only significant residuals at extreme velocities (near $-2.5$ and $-8.5$ km s$^{-1}$ for H$_2$$^{18}$O, and near $-2.0$ and $-9.0$ km s$^{-1}$ for SO$_2$). Based on the line profiles and the observed velocity gradient across the source, we suggest that the velocity structure of VLA3 consists of three components: two exhibiting motions (azimuthal and radial) in the equatorial plane and one static component. 

		\subsubsection{Molecular excitation}	\label{mol_excitation_analysis}
			We estimate the excitation conditions of VLA3 based on the line ratios of the chemically related molecules HDO and H$_2$$^{18}$O. If we assume optically thin emission and local thermodynamic equilibrium (LTE), the total column density at a given excitation temperature can be derived via
			\begin{equation}
			N_{\rm tot} \ = \ \frac{2.04W}{\theta_{a}\theta_{b}}\frac{Q_{\rm tot}e^{E_{\rm u}/T_{\rm ex}}}{g_Ig_KS\mu^2\nu^3}\times10^{20} $\ \ cm$^{-2},  
		\end{equation}
where $W$ is the integrated intensity in Jy beam$^{-1}$ km s$^{-1}$, $\theta_a$ and $\theta_b$ are the FWHM widths of the synthesized beam in arcsec, $Q_{\rm tot}$ is the partition function, $E_{\rm u}$ is the upper energy level in K, $T_{\rm ex}$ is the excitation temperature in K, $g_I$ and $g_K$ are the spin and $K$ degeneracies, respectively, $S$ is the line strength, $\mu$ is the dipole moment in Debye, and $\nu$ is the rest frequency in GHz. The source filling factor is assumed be unity and background emission is neglected. In other words, the excitation temperature can be expressed as a (nonlinear) function of the column density ratio of HDO and H$_2$$^{18}$O, if HDO and H$_2$$^{18}$O share the same excitation temperature. Once the excitation temperature is estimated, the column density can be derived via Equation (3). 

\begin{figure*}[htbp]
	\centering
	\includegraphics[width=17cm]{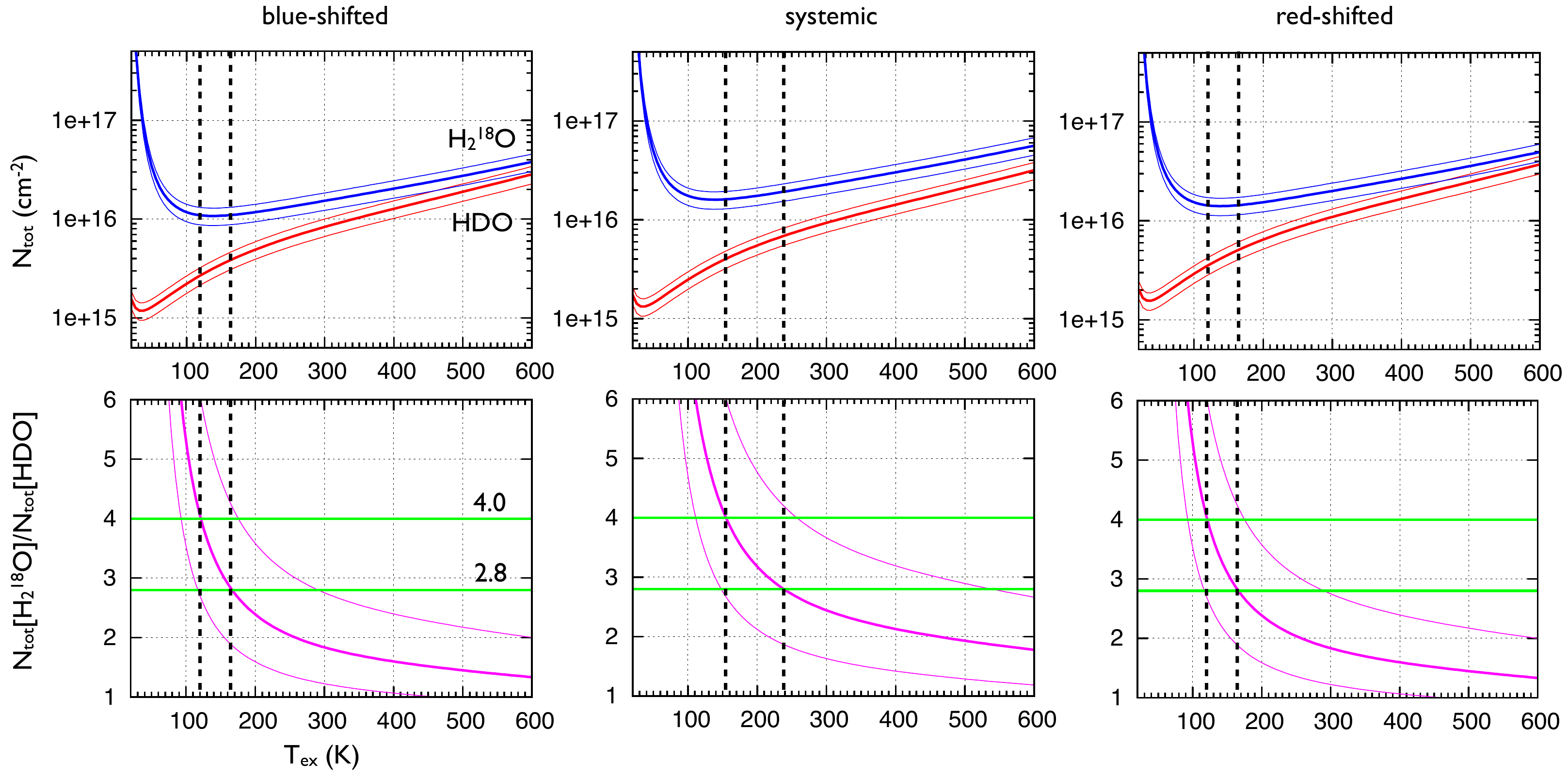}
	\caption{H$_2$$^{18}$O and HDO column density and column density ratio plots. The left, middle and right columns represent the plots for the blue-shifted, systemic and red-shifted components, respectively. In the first row, total column densities at different excitation temperatures are plotted in blue (H$_2$$^{18}$O) and red (HDO) thick curves. Their column density density ratios, shown in the second row, are plotted in thick purple curves. The thin curves represent the errors. In addition, the H$_2$$^{18}$O over HDO column density ratio, derived from \citet{vanderTak06} and \citet{Wilson94}, are plotted as green horizontal lines. The excitation temperatures for a given component can be read out by finding the intersection of the green lines and thick purple lines. The thick vertical dashed lines mark the derived temperatures for ratio 2.8 and 4.0. Here we derive the excitation temperature without considering the error of absolute flux density. Derived temperatures and column densities are summarized in Table \ref{mol_excitation}.}
	\label{fig:h2o_hdo_excit}
\end{figure*}

			The column density ratio $N_{\rm tot}$[H$_2$$^{18}$O]/$N_{\rm tot}$[HDO] can be expressed as 
			\begin{equation}
				\frac{N_{\rm tot}[\rm H_2\,^{18}O]}{N_{\rm tot}[\rm HDO]} = \frac{N_{\rm tot}[\rm H_2O]}{N_{\rm tot}[\rm HDO]}\times\frac{N_{\rm tot}[\rm H_2\,^{18}O]}{N_{\rm tot}[\rm H_2O]}.
			\end{equation}
			For the column density ratio of H$_2$O and HDO, we adopt the abundance ratio [H$_2$O]/[HDO] of 1400--2000 for the inner region of the envelope around VLA3 (assuming this ratio can be applied to the observed rotating structure), which is derived based on the ``jump'' model of 1D radiative transfer calculations \citep[see Table 11 of][]{vanderTak06}. We assume the column density ratio of H$_2$$^{18}$O and H$_2$O to be 0.002 based on the [$^{18}$O]/[$^{16}$O] ratio \citep{Wilson94}. Therefore, the column density ratio $N_{\rm tot}$[H$_2$$^{18}$O]/$N_{\rm tot}$[HDO] is 2.8--4.0. The total column densities of HDO and H$_2$$^{18}$O as a function of $T_{\rm ex}$ are shown in the upper three panels of Fig. \ref{fig:h2o_hdo_excit}, while the column density ratio $N_{\rm tot}$[H$_2$$^{18}$O]/$N_{\rm tot}$[HDO] as a function of $T_{\rm ex}$ can be seen in the bottom three panels of Fig. \ref{fig:h2o_hdo_excit}. From the column density ratio plots, a range of excitation temperatures for a given velocity component can be  estimated by finding the intersections of the derived curves of the column density ratio from the new observations (in purple) and the horizontal lines of the adopted column density ratios from Equation (4) (in green). We summarize the molecular excitation condition of VLA3 in Table \ref{mol_excitation}.
			
			We estimate the excitation temperatures for the blue-shifted, systemic and red-shifted components to be 120--165 K, 155--240 K and 120--165 K, respectively, or 90--290 K, 110--540 K and 90--290 K, respectively, if the error in absolute flux calibration is considered. We find that the blue- and red-shifted components have similar excitation temperatures, implying again  that both components are from a single source structure such as the observed equatorial gas component around VLA3. There is also an indication that the systemic component might be warmer. The overall column densities counting all three velocity components (without absolute error in flux density) for HDO, H$_2$$^{18}$O and SO$_2$ are $1\times10^{16}$, $4\times10^{16}$ and $1\times10^{16}$ cm$^{-2}$, respectively. The fractional abundances are estimated to be $5\times10^{-8}$, $2\times10^{-7}$ and $6\times10^{-8}$ for HDO, H$_2$$^{18}$O and SO$_2$, respectively, adopting the 203.4-GHz continuum peak flux at the resolution of $1\farcs23\times0\farcs72$, P.A. 26$\degr$ and using $T_{\rm d} = 200$ K in the calculation of Equation 1.

			We note that the line ratio analysis of HDO and H$_2$$^{18}$O discussed above is sensitive to the assumed abundance ratios of HDO/H$_2$O and  $^{16}$O/$^{18}$O. For example, if the assumed $N_{\rm tot}$[H$_2$$^{18}$O]/$N_{\rm tot}$[HDO] is smaller (larger) by a factor of two, the corresponding temperatures would be less than 100 K (greater than 250 K) for all three velocity components. Alternatively, if we adopt a temperature of 100 K (the ice evaporation temperature), the derived column density ratio $N_{\rm tot}$[H$_2$$^{18}$O]/$N_{\rm tot}$[HDO] would be less than 7, consistent with \citet{vanderTak06}. Only observations of multiple transitions of HDO and H$_2$$^{18}$O can further constrain the excitation conditions of the inner region of VLA3.
		
\begin{figure}[htbp]
	\centering
	\includegraphics[width=8cm]{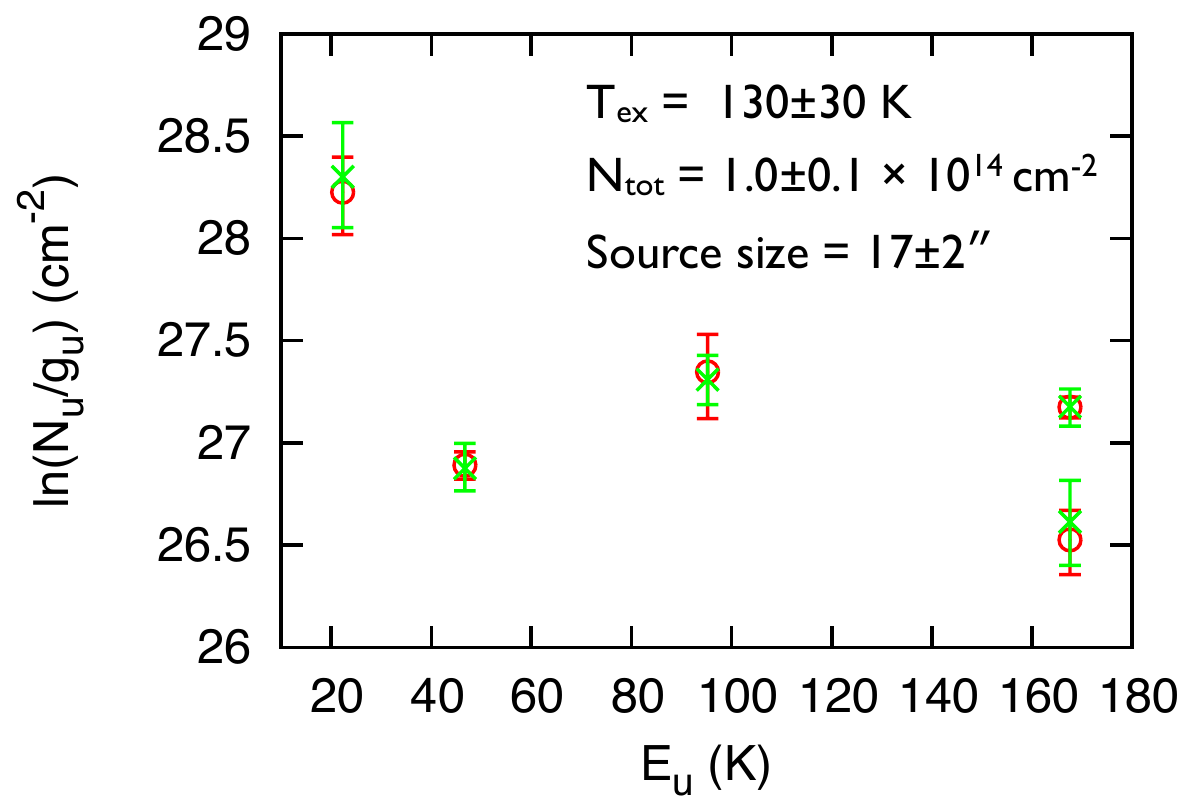}
	\caption{HDO rotation diagram plotted with the data published by \citet{vanderTak06}. Red open circles are the data observed with JCMT and IRAM 30m. The green crosses represent the best-fit model from population diagram analysis \citep{Goldsmith99}.}
	\label{fig:HDO_PDA}
\end{figure}

		 Alternatively, the excitation conditions on large scales near VLA3 can be estimated from the HDO single-dish data published by \citet{vanderTak06} with the population diagram method \citep{Goldsmith99}. Including the effects of beam dilution and line opacity, the excitation temperature, total column density and source size are solved self-consistently. The observed data (red open circles) and best fit model (green crosses) are plotted in Fig. \ref{fig:HDO_PDA}. The excitation temperature is estimated to be $130\pm30$ K with a column density of $\sim(1.0\pm0.1)\times10^{14}$ cm$^{-2}$ distributed uniformly in a source with angular size of $17\pm2\arcsec$. Line opacities of all the transitions used are estimated to be less than 0.006. The excitation temperatures derived from the single-dish data are consistent with the ones derived from our new interferometric data. The total HDO column density including all three velocity components is about $3\times10^{13}$ cm$^{-2}$ if we observe the same structure (roughly $0.8\arcsec-1.0\arcsec$) seen by the interferometer with a $17\arcsec$ beam which is consistent with the value derived from single-dish data if missing flux is considered. As a result, both single-dish and interferometric data imply that HDO is present in both the outer envelope and the inner dense region toward VLA3 with a temperature greater than $\sim$100 K.		 						
\begin{table*}
	\centering
	\caption{Line parameters of HDO, H$_2$$^{18}$O and SO$_2$ derived from fitting of three Gaussian components.}	
	\small{
	\begin{tabular}{lccccccccccc}
		\hline\hline
				&	\multicolumn{3}{c}{Blue-shifted}	& &	\multicolumn{3}{c}{Systemic}	& &	\multicolumn{3}{c}{Red-shifted}	\\
		\cline{2-4} \cline{6-8} \cline{10-12}
				&	$V_{\rm LSR}$\tablefootmark{a}	&	$\Delta V$\tablefootmark{b}	& $W$\tablefootmark{c} &	&	$V_{\rm LSR}$	&	$\Delta V$	&	$W$ & &	$V_{\rm LSR}$	&	$\Delta V$	&	$W$ \\
		Molecule		& (km s$^{-1}$) & (km s$^{-1}$) & (Jy beam$^{-1}$ & & (km s$^{-1}$) & (km s$^{-1}$) & (Jy beam$^{-1}$ &  & (km s$^{-1}$) & (km s$^{-1}$) & (Jy beam$^{-1}$ \\		
		             &               &               &   km s$^{-1}$) & & & &  km s$^{-1}$) & & & & km s$^{-1}$) \\ 
		\hline
		HDO & $-6.89$(3) & 1.20	(5) & 0.072(3) & & $-5.35$(34) & 1.20(5) & 0.080(3) & & $-3.92$(26) & 1.20(5) & 0.094(3) 	\\
		H$_2$$^{18}$O & $-6.89$ & 1.45 & 0.544(19) & & $-5.35$ & 1.45 & 0.807(19) & & $-3.92$ & 1.45 & 0.712(19)   \\
		SO$_2$ & $-6.89$ & 1.45 & 1.64(11) & & $-5.35$ & 1.45 & 1.74(12) & & $-3.92$ & 1.45 & 2.20(11) 	\\
		\hline
	\end{tabular}}					
	\tablefoot{
		Errors in units of the last decimal are given in the parentheses. Values without parentheses are fixed in the fitting.\\
		\tablefoottext{a}{Line center velocity.} \\
		\tablefoottext{b}{FWHM line width.} \\
		\tablefoottext{c}{Integrated intensity.}}	
	\label{tab:lines}
\end{table*}
\begin{table*}
	\centering
	\caption{Excitation temperatures and column densities.}	
	\small{
	\begin{tabular}{lccccccc}
		\hline\hline
		   &     $T_{\rm ex}$\tablefootmark{a}  &  $N_{\rm tot}$[HDO]\tablefootmark{b}  &  $N_{\rm tot}$[H$_2$$^{18}$O]\tablefootmark{b}  &  $N_{\rm tot}$[SO$_2$]\tablefootmark{b} & $X$[HDO]\tablefootmark{c} & $X$[H$_2$$^{18}$O]\tablefootmark{c} & $X$[SO$_2$]\tablefootmark{c}\\
		Component  &  (K)  &  ($\times10^{15}$cm$^{-2}$)  &  ($\times10^{16}$cm$^{-2}$)  &  ($\times10^{15}$cm$^{-2}$) & ($\times10^{-8}$) & ($\times10^{-7}$) & ($\times10^{-8}$) \\		 	
		\hline
		Blue-shifted	&  120--165  &  2.7--3.9   &  1.1  &  3.0--4.8   & ... & ... & ...	\\
		Systemic  	&  155--240  &  4.0--5.5   &  1.6  &  4.7--7.3   & ... & ... & ...	 \\
		Red-shifted	&  120--165  &  3.5--5.1   &  1.4  &  4.0--6.5   & ... & ... & ... \\
		Overall		&  ...   &  10.2--14.5 &  4.1  &  11.7--18.6 & 4.4--6.2 & 1.7 & 4.9--7.9 \\
		\hline
	\end{tabular}
	}					
	\tablefoot{ \\
		\tablefoottext{a}{Excitation temperatures estimated from column density ratios. See Sect. 3.2.4 for details.} \\
		\tablefoottext{b}{Total column density derived under LTE and optically thin assumptions.} \\
		\tablefoottext{c}{With respect to H$_2$. Assuming a dust temperature of 200 K, $N_{\rm tot}$[H$_2$] of $2.35\times10^{23}$ cm$^{-2}$ is derived from the 203.4-GHz continuum at a resolution of $1\farcs23\times0\farcs72$, P.A. 26$\degr$. The uncertainty in $N_{\rm tot}$[H$_2$] is not considered.}}
\label{mol_excitation}
\end{table*}

\section{Kinematics of the equatorial region of VLA3}
	\subsection{Evidence of rotation and outward radial motions}
\begin{figure}
	\centering
	\includegraphics[width=8cm]{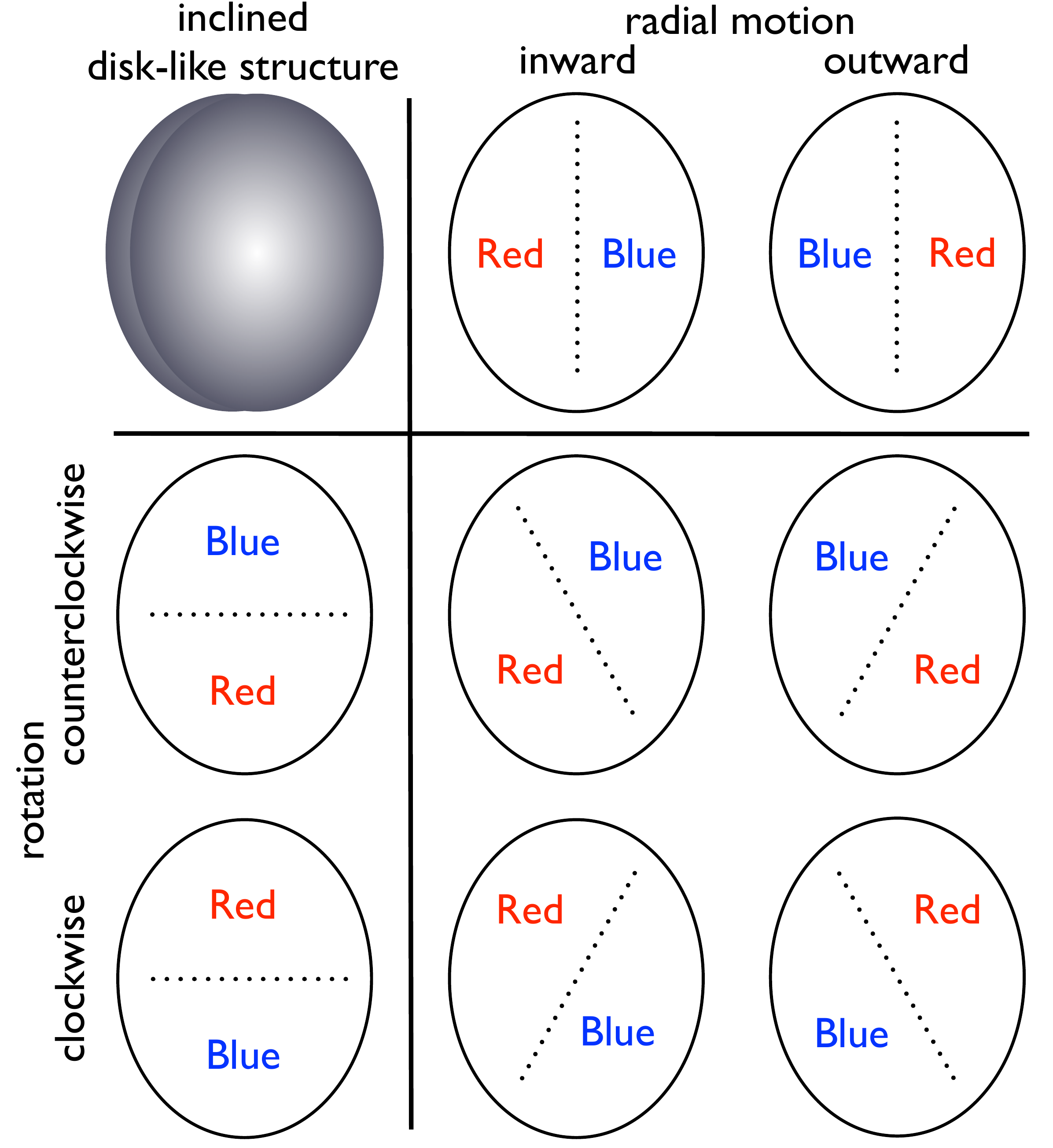}
	\caption{Schematic view of the velocity fields combining radial motion and rotation.}
	\label{fig:vmodel_cartoon}
\end{figure}

		The new (sub)arcsecond resolution PdBI observations resolve the velocity field in the inner region of \object{AFGL 2591--VLA3}. Clear {\it linear} velocity gradients in the northeast-southwest direction are observed in HDO, H$_2$$^{18}$O and SO$_2$ lines. Comparison of the distributions of isovelocity contours of these lines reveals that HDO and H$_2$$^{18}$O best reflect the kinematics of the equatorial region of VLA3, while SO$_2$ is more affected by the east-west outflow. In addition, a triple-peaked line profile is observed in HDO. Given the bipolar outflow geometry in the east-west direction \citep{Mitchell92,Hasegawa95,Preibisch03}, which defines a north-south equatorial plane, we propose that the observed lines trace rotation in the inner envelope around VLA3. We suggest that radial motions alter the velocity field's position angle from the expected north-south direction to the observed northeast-southwest direction. Consider an inclined disk-like structure at P.A. $0\degr$, with  rotation and radial motion but without vertical motion (as expected for a model in vertical hydrostatic equilibrium), four possible velocity patterns exist (as shown schematically in Fig. \ref{fig:vmodel_cartoon}), depending on whether the rotation appears clockwise or counterclockwise as seen by the observer, and whether the radial motion are inward or outward. Only the combination of counterclockwise rotation and {\it outward} radial motions reproduce the observed velocity pattern (c.f. Fig. \ref{line_maps} and \ref{fig:vmodel_cartoon}).
	
	\subsection{A toy model}	
\begin{figure}
	\centering
	\includegraphics[width=8.5cm]{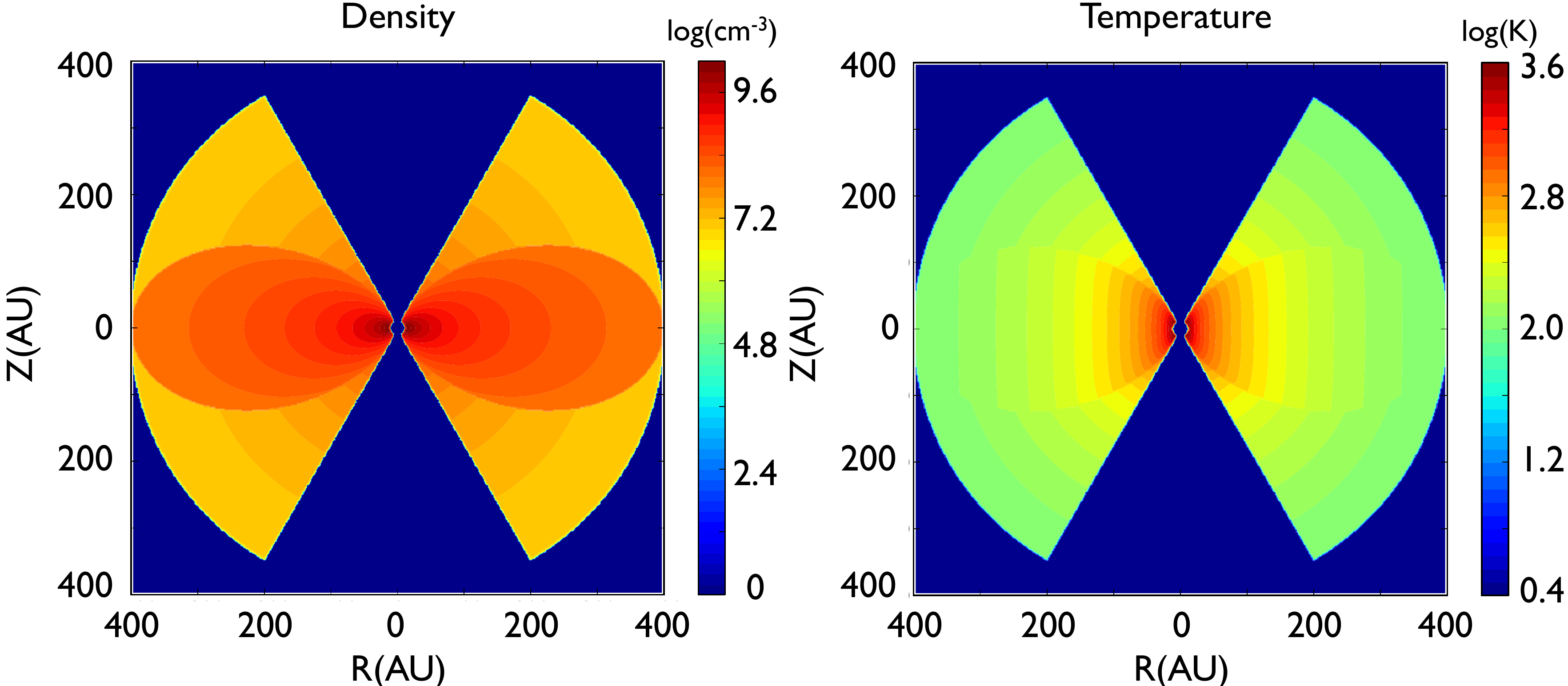}
	\caption{Gas density (left) and gas temperature (right) distributions of the source in the radiative transfer models, including a disk-like structure, a spherical turbulent envelope, and an outflow cavity.}
	\label{fig:model_profile_plot}
\end{figure}

		 To explore if a kinematical model as suggested in the previous section can reproduce the observed velocity pattern and molecular line spectra, we constructed a toy model and calculate the resulting emission using the axisymmetric radiative transfer code {\it RATRAN} \citep{Hogerheijde00}. Our model includes a flattened ``disk-like'' structure surrounded by a static spherical envelope and a bipolar outflow cavity (Fig. \ref{fig:model_profile_plot}). We adopt a parametrized description of the density, temperature, and velocity field, and assume LTE and optically thin conditions for the line emission. Given the critical densities of the observed transitions ($7\times10^{4}$, $3\times10^{5}$ and $5\times10^{6}$ cm$^{-3}$ for HDO, H$_2$$^{18}$O and SO$_2$ at 200 K, respectively), and the densities derived in Sect. 3.1.2, the excitation is likely in LTE. We scale the model intensity to match the observed line strengths, which means that we can derive conclusions about the velocity field but not the density or temperature. Our only aim is to show that our simple toy model can reproduce the observed velocity pattern and spectral line shapes. We do not perform any global model optimization in density, temperature and velocity profiles. Instead, for fixed  sets of density and temperature profiles, we conduct a simplified optimization to the parametrized velocity field. Our derived parameters are therefore not best-fit solutions but rather indicative values. 

\begin{figure}
	\centering
	\includegraphics[width=8cm]{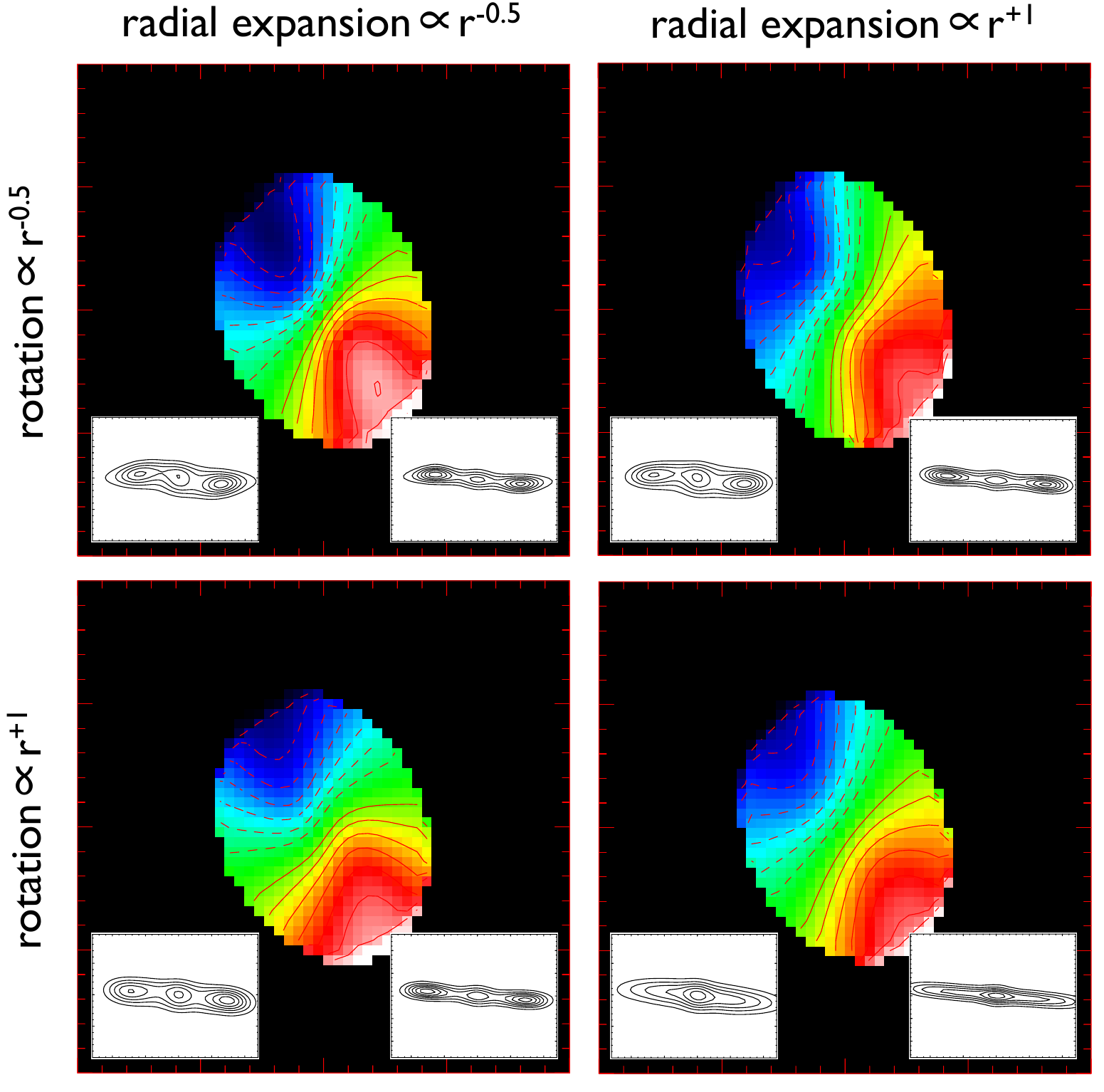}
	\caption{Model moment 1 maps of the combined velocity profiles in azimuthal and radial directions. The model position-velocity maps are included in each panel (bottom-left: P.A. $0\degr$, bottom-right: P.A. $45\degr$). In all cases, linear velocity gradient can be inferred. However, only Keplerian-like rotation ($\sim r^{-0.5}$) plus Hubble-law like radial expansion ($\sim r^{+1}$) can reproduce the observed mirrored S-shape isovelocity contours seen in H$_2$$^{18}$O (c.f., Fig. \ref{line_maps}(e)).}
	\label{vr_va_toy_model}
\end{figure}	
\begin{figure}
	\centering
	\includegraphics[width=8cm]{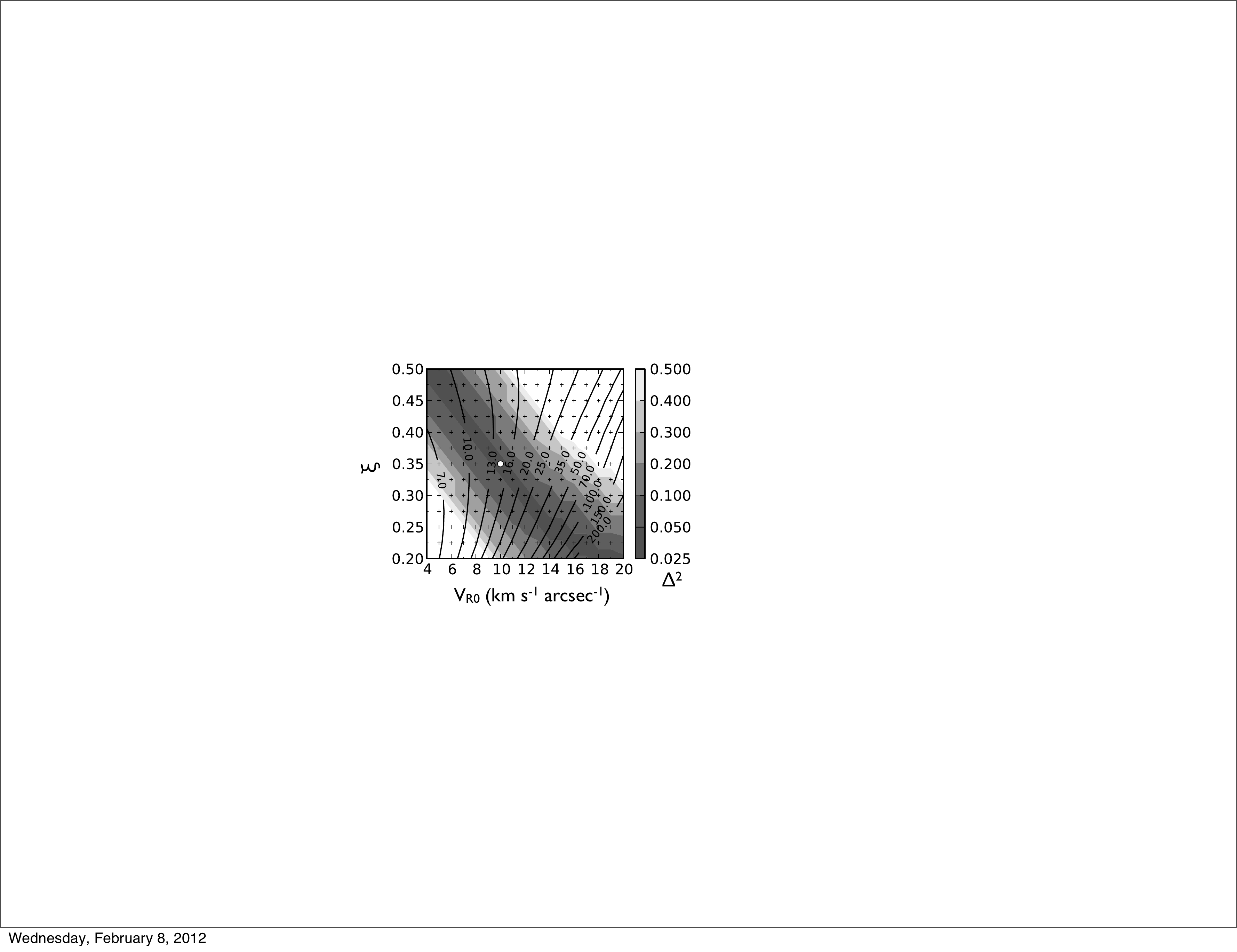}
	\caption{Simplified parameter optimization of our kinematic model. The grey scales represent the similarity of model and observed HDO triple-peaked line profile (lower values for better match). The solid curves with values in unit of km s$^{-1}$ arcsec$^{-1}$ indicate the measured strength of linear velocity gradient at P.A. $0\degr$ in the grid models. The open circle highlights our best matching parameter set ($v_{R0}$ = 10 km s$^{-1}$ arcsec$^{-1}$, $\xi$ = 0.35).}
	\label{hdo_chi2_surface}
\end{figure}	
		\subsection{Details of the model}
			The density profiles for the disk-like structure (for simplicity, we use ``disk'' hereafter in this section) and the envelope are assumed to follow power-law distributions, respectively, as $n_{\rm disk}(r,\theta) = An_{\rm e0}(r/r_{\rm d})^{-\alpha}\sin^{f}(\theta)$ and $n_{\rm env}(r) = n_{\rm e0}(r/r_{\rm d})^{-\alpha}$, where $n_{\rm e0}$ is the density at radius $r_{\rm d}$ (the outer radius of the disk), $A$ is the disk-to-envelope density ratio at $r_{\rm d}$ in the mid-plane, $\theta$ is the angle measured from the polar axis, $f$ is the flattening parameter and $\alpha$ is the power-law index. In the disk density profile, regions with densities lower than $An_{\rm e0}$ are set to have zero density, resulting in a dumpling-like, flattened structure. The overall density in the model is the sum of $n_{\rm disk}$ and $n_{\rm env}$. For the temperature profiles in the disk and the envelope, we assume power-law distributions, respectively, as $T_{\rm disk}(R) = BT_{\rm e0}(R/r_{\rm d})^{-\beta}$ and $T_{\rm env}(r) = T_{\rm e0}(r/r_{\rm d})^{-\beta}$ (i.e., we adopt a vertically isothermal model for the disk) where $R$ denotes the radial direction in cylindrical coordinates, $T_{\rm e0}$ is the temperature at $r_{\rm d}$, $B$ is the disk-to-envelope temperature ratio, and $\beta$ is the power-law index. The gas temperature for a given position is the weighted mean of the disk and envelope values: $T = (n_{\rm disk}T_{\rm disk}+n_{\rm env}T_{\rm env})/(n_{\rm disk}+n_{\rm env})$. The distance to the source is assumed to be 1 kpc. The inner radius of the model is set to 12.5 AU \citep[half the size of the H{\small II} region;][]{vanderTak05} and the outer radius to 400 AU ($=r_{\rm d}$, based on the size derived from our 203.4 GHz continuum). In addition, a bipolar outflow cone with an opening angle of $60\degr$ is included in the model by setting the density inside this cone to be zero. We describe the velocity field in the disk by three orthogonal components as radial ($v_R$), vertical ($v_z$) and azimuthal ($v_a$). The velocity field in the static envelope is set to be purely turbulent, inspired by the observed spectra showing emission peaks near systemic velocity. The gas velocity for a given position is the weighted mean of the disk and envelope values: $\vec{v} = (n_{\rm env}\vec{v_{\rm env}}+n_{\rm disk}\vec{v_{\rm disk}})/(n_{\rm disk}+n_{\rm env})$. The inclination of the source is set to be $30\degr$ \citep{vanderTak06} with the blue-shifted outflow cone pointed to the west. Model parameters are chosen to ensure that the LTE and optically thin assumptions are valid. We adopt $A=10$, $n_{\rm e0}=1\times10^{7}$ cm$^{-3}$, $\alpha=1.5$, $f=5$, $B=1$, $T_{\rm e0}=100$ K, $\beta=1$, a molecular fractional abundance of $1\times10^{-13}$ and a turbulent line width of 0.5 km s$^{-1}$. A detailed description of the parametrized velocity field is given in the next section. We scale the line intensity to the observed values, and only analyze the velocity signatures (moment 1 maps and position-velocity maps) and spectral line profiles. We note that this scaling process makes the exact profiles of density and temperature irrelevant. 
						   		
		\subsection{Evidence of sub-Keplerian rotation and Hubble-law like radial expansion traced by HDO and H$_2$$^{18}$O} 
		 	The analysis of the position-velocity maps of the observed lines toward VLA3 reveals that the velocity field is characterized by a linear velocity gradient (Fig. \ref{fig:pvmaps}), which suggests that the equatorial region is undergoing solid-body rotation. However, the analysis demonstrated in Fig. \ref{fig:vmodel_cartoon} implies that an extra outward radial component is required to reproduce the observed northeast-southwest velocity pattern (Fig. \ref{line_maps} (d)--(f)). Therefore, a direct link between the observed linear velocity gradient and the inferred solid-body rotation may not be trivial. In our toy model, we consider two types of velocity profiles in the azimuthal and radial directions for the disk-like structure: one is proportional to $r^{-0.5}$ and the other follows $r^{+1}$. Although all the combinations of the velocity field can reproduce the triple-peaked line signature seen in HDO and the linear velocity gradient seen in the position-velocity maps, only Keplerian-like rotation ($v_a\sim r^{-0.5}$) plus Hubble-law like radial expansion ($v_R\sim r^{+1}$) can reproduce the mirrored S-shape isovelocity contours observed in H$_2$$^{18}$O, which is less affected by the outflow than SO$_2$ (Fig. \ref{vr_va_toy_model}; c.f., Fig. \ref{line_maps}(e)). An additional velocity component in the $z$ direction may redistribute the isovelocity contours to match the observed SO$_2$ moment 1 map. In our toy model, the outflow component is not included due to the complexity of outflow morphology and velocity field. Therefore, we use HDO and H$_2$$^{18}$O data to constrain the kinematics of the equatorial region of VLA3. To summarize, we adopt the parametrized velocity for the disk-like structure as:
		\begin{equation}
		\begin{array}{l}
		\displaystyle v_R = v_{R0}r\\
		\displaystyle v_a = \xi \sqrt{GM_{\star}/r} \\
		\displaystyle v_z = 0,
		\end{array} 
		\end{equation}	
where $v_{R0}$ is an analog of the Hubble constant, $\xi$ is a constant between 0 and 1, $G$ is the gravitational constant and $M_{\star}$=16$M_{\sun}$ \citep{vanderTak05} is the stellar mass.			
\begin{figure*}[]
	\centering
	\includegraphics[width=17cm]{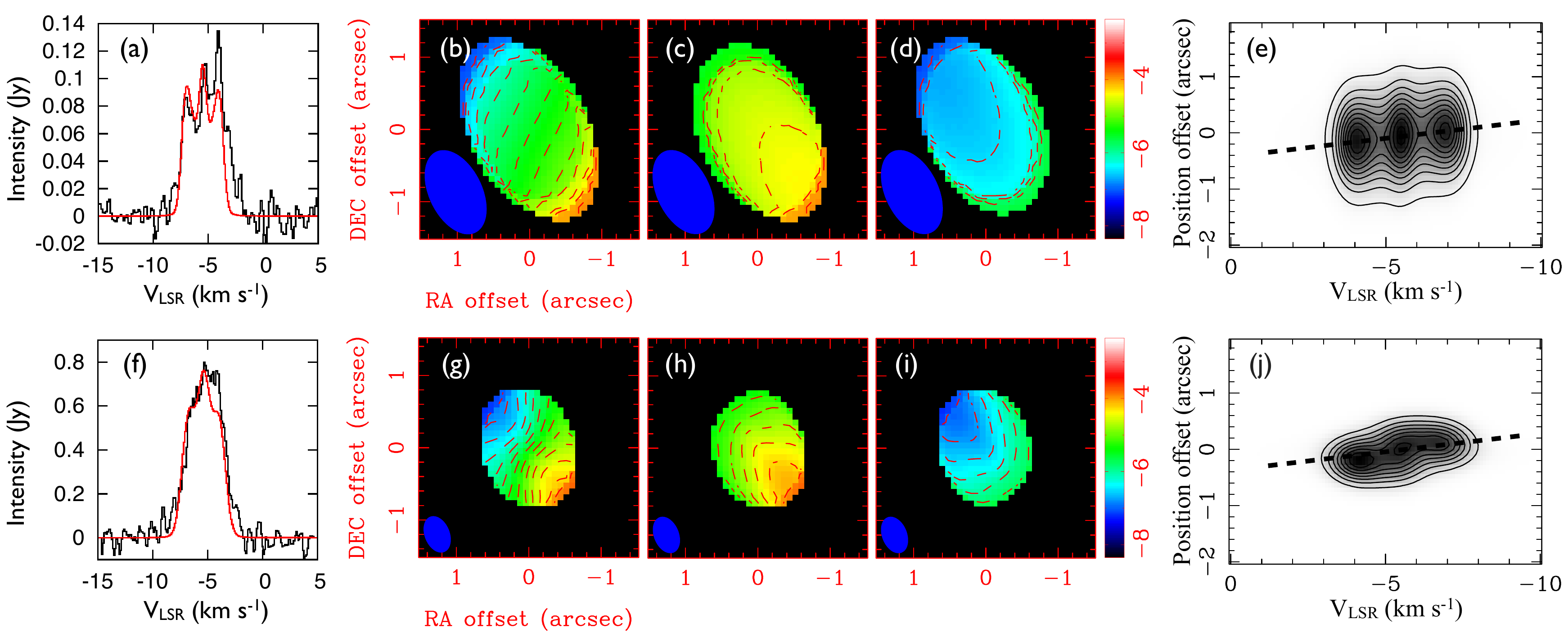}
	\caption{(Left) Best matching model spectrum (a), moment 1 maps (b)--(d) and position-velocity map (f) for HDO. The color scales and contour levels for moment 1 maps are identical to Fig. \ref{line_maps} (d)--(l). The dashed line in the position-velocity map represents a linear velocity gradient with strength of 15 km s$^{-1}$ arcsec$^{-1}$. (Right) Best matching model for H$_2$$^{18}$O. }
	\label{ratran_model_V3}
\end{figure*}	
	 	To estimate the strength of the two free parameters $v_{R0}$ and $\xi$ in our kinematic model, we perform a simplified parameter optimization which utilizes the peak velocities of the HDO triple-peaked line profile ($-6.7$, $-5.4$ and $-3.9$ km s$^{-1}$ for the blue-shifted, systemic and red-shifted components, respectively), and the measured linear velocity gradient of $\sim15$ km s$^{-1}$ arcsec$^{-1}$ at P.A $0\degr$ as the constraints. We constructed a grid of models for HDO with $v_{R0}$ between 4 and 20 km s$^{-1}$ per 1000 AU and $\xi$ between 0.2 and 0.5. For each set of parameters ($v_{R0}$, $\xi$), we compute $\Delta^2 = \Sigma({\rm observed \ peak \ velocities - model \ peak \ velocities})^2$, which is smaller for a better match. Figure \ref{hdo_chi2_surface} shows that the models with the lowest $\Delta^2$  values span a diagonal in ($v_{R0}$, $\xi$)-space, indicating that the model is degenerate. To break the degeneracy, we compute the strength of the linear velocity gradient at P.A. $0\degr$ in our grid model and plot it as the solid curves in Fig. \ref{hdo_chi2_surface}. By matching the observed value, we conclude that based on our toy model, the azimuthal motion of the disk-like structure traced by HDO can be characterized by sub-Keplerian rotation (about 1/3 the strength of Keplerian rotation with a stellar mass of 16 $M_{\sun}$) and the radial motion is the accelerated expansion with the analogue Hubble constant of 10 km s$^{-1}$ per 1000 AU at 1 kpc. The best matching model spectrum, moment 1 maps and position-velocity map for HDO can be found in Fig. \ref{ratran_model_V3} (a)--(e). 
			
		To model the H$_2$$^{18}$O data, we use the best matching solution for HDO ($v_{R0}$ = 10 km s$^{-1}$ arcsec$^{-1}$, $\xi$ = 0.35) as the starting point. Certainly, the turbulent line width of 0.5 km s$^{-1}$ is too narrow for H$_2$$^{18}$O since its line profile is Gaussian-like with hints of secondary velocity components. We therefore adopt a turbulent line width of 1.0 km s$^{-1}$ in the model. The best matching model for H$_2$$^{18}$O can be seen in Fig. \ref{ratran_model_V3} (f)--(j). Though not perfectly matching the observations, our toy model of kinematics is able to reproduce both HDO and H$_2$$^{18}$O data qualitatively. In our toy model, we found that the kinematics of the proposed disk-like structure around VLA3 is best described by two velocity components: sub-Keplerian rotation and Hubble-law like expansion. We also found that the conclusion of rigid-body rotation across the source judging directly from the linear velocity gradient observed in position-velocity maps may not be made if the observed velocity gradient does not align well with the equatorial plane defined by the outflow.

\section{Discussion}
	\subsection{An embedded disk-like structure in \object{AFGL 2591--VLA3}?}
		Our (sub)arcsecond PdBI images of HDO, H$_2$$^{18}$O and SO$_2$ reveal a clear velocity gradient in the northeast-southwest direction across the source \object{AFGL 2591--VLA3} (Sect. 3.2.2). Together with the shape of the HDO line profile (Sect. 3.2.3), and the symmetric temperature distribution of the blue- and red-velocity components (Sect. 3.2.4), we propose that the source \object{AFGL 2591--VLA3} is surrounded by a disk-like structure undergoing rotation and expansion. The kinematics of the equatorial region is further characterized as sub-Keplerian rotation plus Hubble-law like expansion as shown by radiative transfer modeling of a toy model (Sect. 4). In the models for HDO, a microturbulent line width of $\sim0.5$ km s$^{-1}$ matches the triple-peaked line profile, which appears small for the violent environment of massive star-forming regions \citep{Zinnecker07}. On the other hand, a microturbulent line width of 1.0 km s$^{-1}$ is needed to match the almost single-peaked line profiles seen in H$_2$$^{18}$O. We expect that an even larger turbulent line width is required to model SO$_2$, judging from the observed SO$_2$ spectrum. This suggests that these molecules trace different regions within the disk-like structure. 
		
\begin{figure}
	\centering
	\includegraphics[width=9cm]{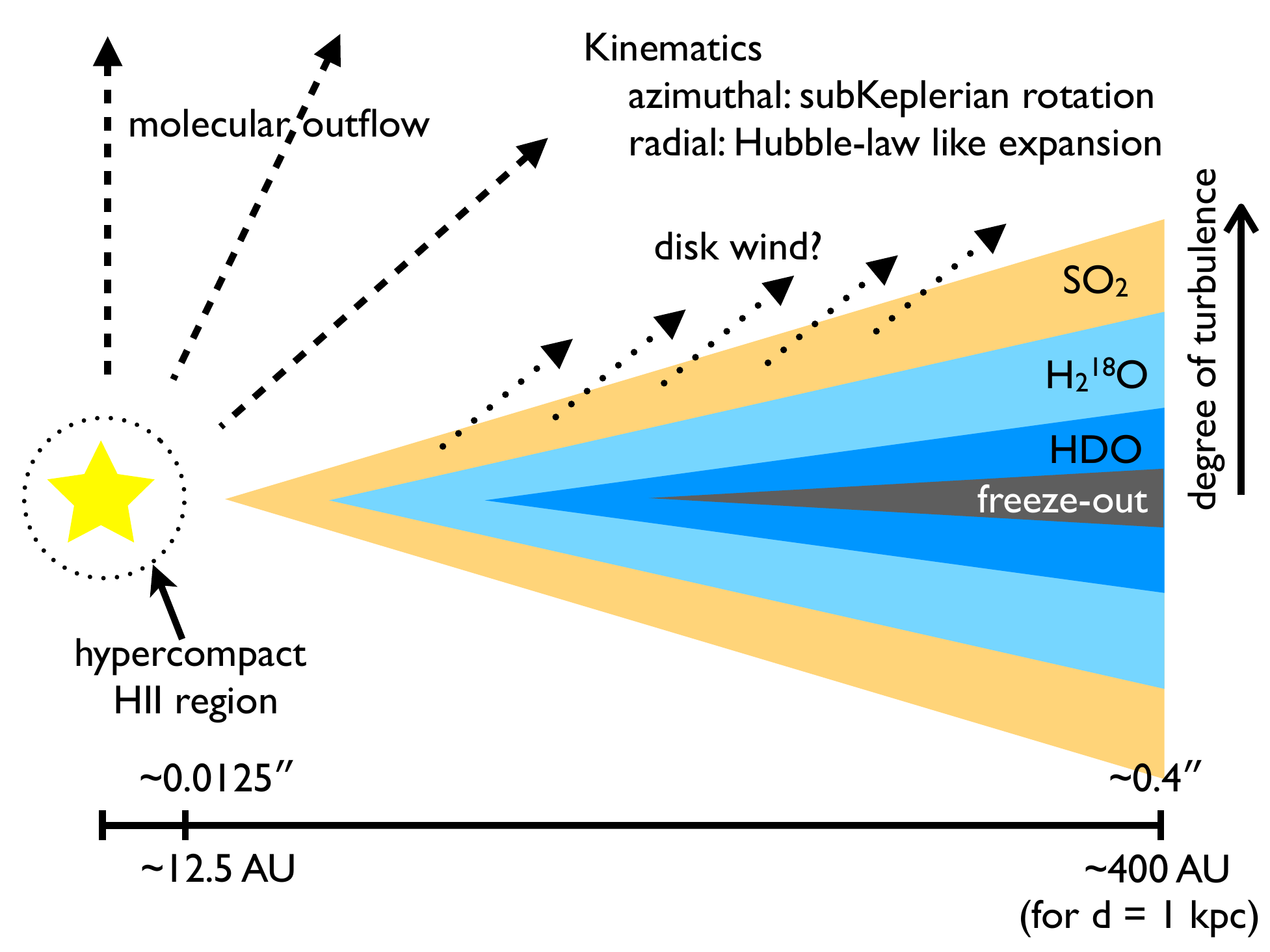}
	\caption{A cartoon showing the idea of a layered disk-like structure around \object{AFGL 2591--VLA3}. HDO is close to the cooler, quiescent mid-plane. H$_2$$^{18}$O is distributed in the warm mid-layer and SO$_2$ originates in the upper disk layers.}
	\label{fig:layered_structure}
\end{figure}

		We hypothesize that the disk-like structure has a layered distribution with HDO close to the mid-plane, H$_2$$^{18}$O in the warm mid-layer and SO$_2$ in the upper disk layers (Fig. \ref{fig:layered_structure}). In addition, significant emission from all three molecules originates in a static turbulent envelope surrounding the disk-like structure, especially for H$_2$$^{18}$O and SO$_2$ where it dominates the line profiles (Table \ref{mol_excitation}). In the disk-like structure, we find temperatures of about 120--165 K, well above the evaporation temperature of water ice \citep{Fraser01}. The HDO molecules may be formed on grain surfaces \citep{Tielens83, Parise05}, and released into the gas phase when the disk-like structure is heated by shocks, stellar radiation and winds. Compared to HDO, H$_2$$^{18}$O originates from a more turbulent region at larger height with increased turbulence in the disk-like structure. In this picture, the presence of H$_2$$^{18}$O but absence of HDO in the mid-layer of the disk-like structure could be due to increased local water formation because of warm gas chemistry as opposed to evaporation of deuterated ices formed at low temperature. SO$_2$ is a good shock/outflow tracer in massive star-forming regions \citep{Schilke97}. Judging from the facts that the observed SO$_2$ line profile has significant emission from line wings, and that a relatively large velocity gradient is measured in SO$_2$ compared to HDO/H$_2$$^{18}$O, we suggest that SO$_2$ traces a part of the disk-like structure where the outflow interacts with the upper layers. The concentric velocity patterns seen both in the blue- and red-shifted part of moment 1 maps of SO$_2$ (Fig. \ref{line_maps}(i) and (l)) likely support the idea since the most extreme velocities occur in the region close to the star (unlike HDO and H$_2$$^{18}$O) as expected for an outflow origin.
		
\subsection{Kinematics of the proposed disk-like structure and comparison to other disk sources}
	 Our toy model suggests that the proposed disk-like structure around VLA3 undergoes sub-Keplerian rotation which cannot be directly inferred from the observed position-velocity maps. We speculate that a magnetic field slows the rotation down below pure Keplerian, which may help to transport the angular momentum outward in VLA3. Similar to the formation of low-mass stars, the magnetic field has been suggested to play an important role in the formation of high-mass stars as well \citep{Girart09,Beuther10}. In strongly magnetized clouds during collapse, magnetic braking is thought to regulate the formation of disks around stars \citep[e.g.,][]{Basu94,Allen03,Mellon08}. Certainly in our case, the magnetic field does not fully dominate the kinematics of VLA3, otherwise a solid-body like velocity field would be expected in the inner region of the disk-like structure \citep[c.f.][Fig. 2]{Galli06}. From OH maser observations, \citet{Hutawarakorn05} measure a maser disk with radius about 750 AU and the magnetic field strength of a few mG toward \object{AFGL 2591--VLA3}, which may be sufficient to slow down the rotation. Subarcsecond dust polarization measurements of the magnetic field toward VLA3 and hence deriving the mass-to-flux ratio would be helpful to test our hypothesis if the magnetic field indeed regulates the kinematics of the equatorial regions of young high-mass stars.
	 
	 In addition to sub-Keplerian rotation, we found that Hubble-law like expansion occurs in the equatorial region of VLA3. The overall kinematics is similar to the SiO maser disk around \object{Orion--IRc2} in which Keplerian rotation and decelerated expansion characterize the kinematics \citep{Plambeck90}. The expanding motions in the disk could be due to the interaction between the stellar radiation and winds with the disk-like structure, providing the forces to push the gas outward \citep[e.g.,][]{Hollenbach94,Yorke96,Richling97}. On the other hand, the presence of an expanding velocity component in the disk-like structure could be the result of accretion through the mid-plane of the very inner region so that the expansion helps to conserve the angular momentum in the equatorial region. The infrared speckle imaging of \object{AFGL 2591} by \citet{Preibisch03} suggests that the outflow or wind is variable in time thus the accretion in the disk may not be continuous but episodic. In addition, little free-free emission is detected toward the source, indicating an evolutionary stage before the development of an ultracompact H{\small II} region \citep{vanderTak05}. As a result, the main accretion phase might have ceased although models show that massive young stars can continue to gain mass through the disk even after an ultracompact H{\small II} region has been  developed \citep[e.g.,][]{Keto07}. Future high-resolution observations of multiple HDO lines and other dense gas tracers, such as N$_2$D$^{+}$ and CH$_3$CN, would be useful to constrain the physical condition and the kinematics of the disk-like structure around VLA3 better and understand its evolutionary stage of high-mass star formation. 
	 
	 Although massive disks showing pure Keplerian rotation traced by various molecules are reported such as the cases in \object{IRAS 20126+4104} \citep[C$^{34}$S;][]{Cesaroni05}, \object{AFGL 490} \citep[C$^{17}$O;][]{Schreyer06} and \object{NGC 7538S} \citep[H$^{13}$CN;][]{Sandell03}, cases showing non-Keplerian rotation were observed as well such as \object{AFGL 2591} (HDO, H$_2$$^{18}$O; this work) and \object{IRAS 18089$-$1732} \citep[NH$_3$;][]{Beuther08}. In addition, massive disk candidates have a wide range of mass and size distributions derived via various tracers as summarized by \citet{Cesaroni07}. Despite the fact that the detections of massive disks to date are still limited, Keplerian rotation in the disks is typically found toward young massive stars with masses about 10 $M_{\sun}$ such as the cases quoted above. For stellar masses about 16 $M_{\sun}$, massive disks start to show non-Keplerian motion such as rigid-body rotation reported in the literature. For even higher stellar masses, the rotating structures (toroids) typically show solid-body rotation and are gravitationally unstable \citep{Beltran05,Furuya08}. The sub-Keplerian rotation discovered in our work suggests that \object{AFGL 2591--VLA3} may be a special case linking transition of velocity field of massive disks from pure Keplerian rotation to solid-body rotation though definitely more new detections of circumstellar disks around high-mass YSOs are required to examine this hypothesis.
		
\subsection{Effect of the uncertainty in distance}
	The distance to \object{AFGL 2591} is uncertain as discussed by \citet{vanderTak99} and \citet{Rygl11}. In this paper, we assume a distance of 1 kpc with total luminosity of 20000 $L_{\sun}$. From the radiative transfer modeling of HDO/H$_2$$^{18}$O with our toy model, we found that the azimuthal motion of the proposed disk-like structure is dominated by sub-Keplerian rotation which is about 1/3 in strength of pure Keplerian rotation with a stellar mass of 16 $M_{\sun}$, and the radial motion is consistent with Hubble-law like expansion with the analogue Hubble constant of 10 km s$^{-1}$ arcsec$^{-1}$. For other distances, our results remain unchanged except for the relative strengths of rotation and expansion. In particular, $v_{R0}$ scales as $1/d$ and $\xi$ scales as $1/\sqrt{d}$, so that for $d=0.5$ and 3.0 kpc, $v_{R0}$ is changed to 20 and 3.3 km s$^{-1}$ arcsec$^{-1}$, while $\xi$ becomes 0.5 (effective Keplerian mass = 4 $M_{\sun}$) and 0.2 (effective Keplerian mass = 0.7 $M_{\sun}$). We suggest that the disk-like structure cannot undergo pure Keplerian rotation at any of these distances since the effective Keplerian masses are not consistent with the scaled total luminosities. 
	 
\section{Conclusions}
	Our (sub)arcsecond resolution PdBI observations in millimeter continuum and in lines of HDO 1$_{1,0}$--1$_{1,1}$, H$_2$$^{18}$O $3_{1,3}$--$2_{2,0}$ and SO$_2$ $12_{0,12}$--$11_{1,11}$ successfully resolve the inner thousand AU region of the young high-mass star \object{AFGL 2591--VLA3}. We summarize our findings below. 
\begin{enumerate}
  \item From the visibility analysis of millimeter continuum and molecular lines, a compact source with diameter of $\sim800$ AU at 1 kpc is found toward VLA3. Depending on the assumptions, the H$_2$ column density, number density and mass are estimated to be $1\times10^{24}$ cm$^{-2}$, $1\times10^{8}$ cm$^{-3}$ and 0.2 $M_{\sun}$, respectively. 
  \item A linear velocity gradient in the northeast-southwest direction is found across the source, which has an offset in P.A. by $\sim40\degr$ to the equatorial region (P.A. $0\degr$) defined by the east-west outflow. We interpret this kinematical signature as a combination of rotation and radial expansion in the equatorial region of VLA3. 
  \item By introducing a toy model with simplified parameter optimization, we further constrain the rotation to be sub-Keplerian and the radial expansion to be Hubble-law like. This result is independent to the distance of the source. 
  \item We speculate that sub-Keplerian rotation in the equatorial region could be effect of magnetic field which turns the velocity profile from pure Keplerian to sub-Keplerian. The expansion could be due to the interaction between the stellar radiation/winds and the gas in the equatorial region of VLA3.
  \item Based on the molecular line profiles, the chemical structure of the proposed disk appears layered, with HDO near the mid-plane, H$_2$$^{18}$O in the mid-layer and SO$_2$ in the upper layers, judging from their turbulent line widths.
\end{enumerate}

\begin{acknowledgements}
	The authors appreciate the anonymous referee for the critical comments which help the preparation of the manuscript. The authors are also grateful to the staff of the Plateau de Bure telescope for assisting with the observations, especially Jan Martin Winters at IRAM Grenoble. The research of K.-S.W. at Leiden Observatory is supported through a Nederlandse Onderzoekschool Voor Astronomie (NOVA) Ph.D. grant.
\end{acknowledgements}

\bibliographystyle{aa} 
\bibliography{reference.bib} 

\end{document}